# Flow-deformed conformations of entangled polymers as persistent random walks

Ismael Yacoubou-Djima and Yitzhak Shnidman*

*Department of Engineering Science and Physics, College of Staten Island,*

*City University of New York, Staten Island, NY 10314, USA*

August 20, 2007

---

*Corresponding author. Electronic address: shnidman@mail.csi.cuny.edu

ABSTRACT

Evolving structure and rheology across Kuhn scale interfaces in entangled polymer fluids under flow play a prominent role in processing of manufactured plastics, and have numerous other applications. Quantitative tracking of chain conformation statistics on the Kuhn scale is essential for developing computational models of such phenomena. For this purpose, we formulate here a two-scale/two-mode model of entangled polymer chains under flow. Each chain is partitioned by successive entanglements into strands that are in one of two modes: entangled or dangling. On the strand scale, conformation statistics of ideal (non-interacting) strands follows a differential evolution equation for the second moment of its end-to-end distance. The latter regulates persistent random walks sampling conformation statistics of ideal entangled strands on the Kuhn scale, as follows from a generalized Green-Kubo relation and the Maximum Entropy Principle. We test it numerically for a range of deformation rates at the start-up of simple elongational and shear flows. A self-consistent potential, representing segmental interactions, modifies strand conformation statistics on the Kuhn scale, as it renormalizes the parameters controlling the persistent random walk. The generalized Green-Kubo relation is then inverted to determine how the second moment of strand's end-to-end distance is changed by the self-consistent potential. This allows us to devise a two-scale propagation scheme for the statistical weights of subchains of the entangled chain. The latter is used to calculate local volume fractions for each chemical type of Kuhn segments in entangled chains, thus determining the self-consistent potential.

# I. INTRODUCTION

## A. Motivation and Perspective

The evolution of interfacial structure and rheology in inhomogeneous polymer melts and blends during processing strongly affects the physical properties of manufactured plastics. In such fluids, interactions between chemically distinct Kuhn segments drive segregation into domains that are characterized by different segmental composition. The typical width of the interfacial regions between the domains is only a few Kuhn lengths. The statistics of chain conformations affects, and is strongly modified by, flow deformations, as well as by the packing and composition variations of Kuhn segments across the interfaces. Thus quantitative tracking of chain conformation statistics across interfaces is essential for developing computational models of interfacial phenomena in polymer fluids under flow.

In a dynamic self-consistent field (DSCF) theory for unentangled chains [1], this task has been accomplished by using Wiener random walks on a lattice, with anisotropic stepping probabilities controlled by local velocity gradients, and renormalized by the self-consistent interaction potential. Each step in a random walk on the face-centered cubic (fcc) lattice represents a rigid link between a pair of adjacent Kuhn segments. Note that the directions of successive steps in a Wiener random walk are uncorrelated. Chain entanglements make the task more challenging, since they partition the chains into strands that are stretched and oriented by the flow. This imposes a secondary structure on chain conformations at an intermediate (strand) lengthscale (between the chain and the Kuhn scales), and forces correlations between orientations of successive links connecting successive pairs of Kuhn segments along the strand. The main objective of this work is to capture these effects in a model based on the formalism of persistent (correlated) random walks on the fcc lattice.

Our model is a natural extension of a class of models known in the polymer literature as the Self-Consistent Field (SCF) theory [2,3]. At the root of this approach is the Flory theorem: a linear chain at thermodynamic equilibrium in a homogeneous homopolymer melt is effectively ideal (non-interacting) [4]. Such an ideal chain can be modeled by a freely jointed chain of Kuhn segments. Its conformation statistics can be sampled by Wiener random walks either on- or off-lattice, with stepping probabilities that are isotropic with respect to the current stepping directions, as well as uncorrelated with the previous stepping directions. The propagator for chain conformations on-lattice (off-lattice) is governed by a master (Chapman-Kolmogorov) equation, assuming the form of a diffusion equation with an isotropic Gaussian solution in the continuum limit [5].

Historically, the domain of applicability of the SCF approach expanded in stages, along the lines of increasing complexity of polymer fluids that it aims to model: from homogeneous homopolymer melts at thermodynamic equilibrium, to inhomogeneous, macro- or micro-phase-separated homopolymer blends and block copolymers deforming under flow. The key for unlocking the next level of complexity is to understand how it modifies the segmental and conformation statistics of a linear chain that is immersed in the fluid.

In inhomogeneous fluids, a Kuhn segment's interaction with chemically distinct adjacent segments and walls can be approximated by a self-consistent potential [6-10]. At sufficient strength, it may drive macro- and/or micro-phase separation into domains with distinct composition, separated by interfacial regions that are typically only a few Kuhn lengths wide. Strong variation of the self-consistent potential across interfaces drastically modifies segmental and conformation statistics of chains in interfacial regions. It renormalizes the stepping probabilities in the Wiener random walk model for ideal chains by a multiplicative factor, corresponding to the statistical weight of placing a free (unconnected) Kuhn segment of a specified species, thus accounting for its interaction with the self-consistent potential at the destination site. The renormalized stepping probabilities are anisotropic across interfacial regions, and the master (Chapman-Kolmogorov) equation governing the renormalized chain propagator turns into an equation for diffusion in a potential field in the continuum limit. If the fluid is at thermodynamic equilibrium, the statistical weight of the free segment is related to the canonical probability to find a free segment at that site in the local mean-field approximation.

Conformation statistics of both ideal and interacting chains are also modified by flow deformations driving the system out of thermodynamic equilibrium. In the DSCF theory for inhomogeneous polymer fluids in the unentangled regime [1], this effect was captured by imposing time-dependent anisotropic stepping probabilities in the Wiener random walk on the Kuhn-scale fcc lattice representing ideal chains deformed by the flow. The anisotropic stepping probabilities were obtained from a system of linear equations relating them to the second moment of the chain's end-to-end distance tensor, which evolved according to a FENE-P dumbbell model in continuum space [11]. As in equilibrium, interactions with chemically distinct segments and walls are represented by a self-consistent potential that drives macro- and/or micro- phase separation, but also renormalizes the anisotropic, time-dependent stepping probabilities by a multiplicative factor, and thus affects the intrachain correlations and chain conformations across interfaces. This factor still corresponds to the statistical weight for placing a free (unconnected) segment at the destination site. However, when the system is out of thermodynamic equilibrium, the free segment probability may depend on time, and is determined from probabilistic transport equations for local segmental probabilities and momentum that are coupled to the FENE-P dumbbell evolution equation. Comparison of DSCF predictions with molecular dynamics simulations of chain conformations (as well as of composition profiles and slip velocities) at interfaces of sheared immiscible, *unentangled* blends showed good semi-quantitative agreement with the results of molecular dynamics simulations [12].

When the number of Kuhn segments in each chain of an equilibrated homogeneous melt consisting of linear homopolymers exceeds a certain threshold, the chains are entangled. In an entangled chain, the number of Kuhn segments between successive entanglements fluctuates about the mean number $N_{\mathrm{e}}$. In molecular dynamics simulations of bead-spring models [13,14], the mean number of beads between successive entanglements in a homogeneous equilibrated melt of entangled chains was estimated in the range between 28 and 80, depending on the chain length and on the method of estimation. Since each Kuhn segment is equivalent to $\approx 1.3$ beads in these models, this gives an estimate of $22 \le N_{\mathrm{e}} \le 62$. For simplicity, we assume for the remainder of this paper $N_{\mathrm{e}} = 50$ Kuhn segments. According to the Doi-Edwards theory [15], chain *strands* between successive entanglements (slip-links) are confined into sections of a tube by an effective potential centered about linear segments of the primitive path (PP).

At thermodynamic equilibrium, conformations of entangled chains have an isotropic Gaussian distribution that is self-similar. The latter means that the conformations of shorter subchains of the entire chain also have an isotropic Gaussian distribution, all the way down to the Kuhn scale. In particular, the conformations of strands between successive entanglements have an isotropic Gaussian distribution as well. Hence, at equilibrium, the length of the PP segment spanning the entangled strand is equal to the diameter of the confining tube, and the orientations of the PP segments along the tube have an isotropic distribution. Flow deformations rotate PP segments toward a preferred orientation, as well as stretch the entangled strands along the preferred orientation, and shrink them along the transverse directions.

Stretching relaxes at the fast Rouse relaxation time, while orientation of PP segments relaxes at the much slower disengagement relaxation time, when deformation rates are low. Hence, approximating a steady but low deformation rate by a sequence of step deformations followed by short relaxation periods, the PP segment length and the tube diameter relax close to their equilibrium values in such a small period, while the orientations of the PP segments do not have sufficient time to relax all the way toward the equilibrium isotropic distribution. However, when deformation rates approach or exceed the inverse of the Rouse relaxation time, PP segment length does not have enough time to relax either, so the strand remains stretched along the principal axis that is parallel to the PP segment, and shrunk along the transverse principal axes. Thus a new length scale emerges when ideal entangled chains are deformed by flow. This length scale corresponds to the (stretched) length of the PP segment between entanglements, and it is larger than the Kuhn length, but smaller than the rms end-to-end distance for the entire chain. The emerging structure on the intermediate scale of the entangled strand breaks the self-similarity of Gaussian conformation statistics that is prevalent at equilibrium.

In a previous attempt at a dynamic SCF model for *entangled* polymer fluids under flow [16], the entangled chain has been modeled as a chain of PP segments. In a continuum limit, such a chain assumes the form of a space curve parameterized by a continuous contour length variable, and the local PP segment orientation along the chain is characterized by the local tangent vector to the curve. The probability distribution over the orientations of the tangent vectors is the continuous analog of the probabilities to step along discrete directions (with Kuhn length steps) in the lattice DSCF model for unentangled chains in Ref. [1]. A recursive Chapman-Kolmogorov equation in Ref. [16] (the continuous analog of the master equation in the discrete DSCF mode for unentangled chains in Ref. [1]) relates the statistical weight of a longer PP subchain to the statistical weight of a shorter PP subchain, accounting for interaction with a self-consistent potential. A second-order expansion of the Chapman-Kolmogorov equation in Ref. [16] results in a differential propagation equation of the diffusion type, (a similar diffusive propagation equation obtained by a second-order expansion of the master equation for the Wiener random walk in [1]). The anisotropic diffusion coefficient in the diffusive propagation equation in [16] corresponds to the second moment of the orientational distribution over the tangent vectors, which is controlled by an integro-differential time evolution equation accounting for advection, flow deformation, reptation and constraint release. In Ref. [1], the anisotropic stepping probabilities of the Wiener random walk on the Kuhn scale fcc lattice, and the anisotropic diffusion coefficient of the diffusive propagation equation are similarly related to the second moment of the end-to-end distance distribution in a FENE-P dumbbell model that evolves in time.

Thus the diffusive propagation in Ref. [16] can be viewed as the continuum limit of a Wiener random walk with anisotropic stepping probabilities. However, in this case the lower bound for the step size is set by the equilibrium length of the PP segments spanning the entangled strands, which we view as a significant limitation, since interfacial structure and dynamics in inhomogeneous entangled melts depend on *short-range* interactions between adjacent Kuhn segments. Hence, an adequate model of dynamic interfacial phenomena in entangled polymer fluids must be sensitive to conformation statistics of strands all the way down to the Kuhn length scale, as it is modified by the combined effects of strand deformation under flow and Kuhn-scale segmental interactions. Our main objective here is to construct a simple version of such a model for flow-deformed conformations of an entangled chain in an inhomogeneous polymer fluid. We named it the two-scale/two-mode (2S2M) model. Some definitions and a schematic description are given below, making our name choice obvious. The mathematical formulation of the model, and some preliminary results of it application to simple flows, are the subjects of the next three sections.

**B. Two-Scale/Two-Mode Model of Entangled Polymer Chains**

Consider a linear homopolymer chain in an entangled polymer melt, consisting of $N^{\alpha}$ freely jointed Kuhn segments of chemical type $\alpha$ . We partition the chain into $n_{\mathrm{e}}^{\alpha}$ entangled strands, spanned between successive entanglements, and a dangling strand at either end (see Fig. 1). We use the variable $k$ as a sequential index for the strands along the chain, starting from a dangling strand. Thus $k=1$ and $k=n_{\mathrm{e}}^{\alpha}+2$ correspond to dangling strands, while $2 \le k \le n_{\mathrm{e}}^{\alpha}+1$ denotes an entangled strand. Experimentally, physical properties of entangled strands exhibit a weak dependence on $k$ . Here we use a two-mode model of the entangled chain that neglects this dependence, as well as any differences between the two dangling strands. Let $N_{\mathrm{e}}^{\alpha}$ and $N_{\mathrm{d}}^{\alpha}$ be the mean number of Kuhn segments in an entangled strand and a dangling strand, respectively. We make the simplifying assumptions that each entanglement (slip-link) is localized about a single

Kuhn segment, and that their positions along the chain are fixed at regular intervals. Hence, in our model, the number of Kuhn segments is fixed at $N_{\mathrm{e}}^{\alpha}$ in an entangled strand, and at $N_{\mathrm{d}}^{\alpha}$ in a dangling strand. In reality, each entanglement is not localized about a single Kuhn segments, and their positions along the chain, as well as the number of Kuhn segments in each strand, are not fixed, but rather fluctuate about the mean values, since Kuhn segments can be exchanged between strands sharing the same entanglement (slip-link).

We identify the last Kuhn segment in strand $k$ with the first Kuhn segment in strand $k+1$, since this is the Kuhn segment at the entanglement (slip-link) shared by the two strands. Correcting for double counting of the shared Kuhn segments at the entanglement points, the total number of Kuhn segment in an entangled homopolymer chain of type-$\alpha$ Kuhn segments is $N^{\alpha} = 2N_{\mathrm{d}}^{\alpha} + n_{\mathrm{e}}^{\alpha}\left(N_{\mathrm{e}}^{\alpha} - 1\right)$. In addition, we assume that for any homopolymer consisting of type-$\alpha$ Kuhn segments, $N_{\mathrm{e}}^{\alpha} = N_{\mathrm{e}} = 50$, and $\frac{1}{2}N_{\mathrm{e}}^{\alpha} \le N_{\mathrm{d}}^{\alpha} \le N_{\mathrm{e}}^{\alpha}$.

We introduce a second variable $s$ to index the sequential number of Kuhn segments between the first and the last Kuhn segments within each strand. In the case of a dangling strand, $s = 1,\ldots,N_{\mathrm{d}}^{\alpha}$, with $s = 1$ always corresponding to its singly-connected segment. In the case of the $k^{\text{th}}$ entangled strand, $s = 1,\ldots,N_{\mathrm{e}}^{\alpha}$ where $s = 1$ labels the segment that is connected to a segment in the $(k-1)^{\text{th}}$ strand, and $s = N_{\mathrm{e}}^{\alpha}$ labels the segment that is connected to a segment in the $(k+1)^{\text{th}}$ strand.

The variables $k$ and $s$ are used here to locate the position of a Kuhn segment along the chain contour on two scales. The value of $k$ locates the strand to which it belongs, thus specifying its position along the chain contour on the coarser scale. The value of $s$ refines the position along the chain contour with Kuhn scale resolution, by specifying the sequential order of the Kuhn segment along this particular strand. In random walk (and differential) models for chain conformations, contour length variables play a *mathematical* role very similar to the role of time in analogous models of motion. For this reason, the contour length variable is sometimes even referred to as "time", though it is dimensionless, and its *physical* role is very distinct from real time, which is denoted here by the variable $t$. This can be very confusing when trying to describe the evolution of chain conformations statistics under flow, since they are characterized by functions of *both* the contour length variable *and* the time $t$. To avoid confusion, while maintaining brevity and the mathematical association with time, we will use the term *sime* for the contour length variable that is specified by the values of $k$ and $s$. Similarly, we will use the terms $s$-displacement ($k$-displacement) and $s$-propagation ($k$-propagation) when describing displacement and propagation in sime on the Kuhn (strand) scale.

The rest of this paper is organized as follows. In Section II we study the conformation statistics of an entangled ideal (non-interacting) strand under flow. On the coarser scale, the time evolution of the second moment of the strand's end-to-end distance is modeled the Marrucci-Ianniruberto differential equation [17] in continuum space. This equation couples affine deformation by the flow with entropy-driven relaxation of strand stretching and orientation on distinct time scales. We test it numerically for a range of deformation rates at the start-up of simple elongational and shear flows. To sample conformation statistics of the strand on the Kuhn scale, we introduce a persistent random walk on the fcc lattice. Section III combines a generalized Green-Kubo relation with the Maximum Entropy Principle to show how the evolving second moment of the strand's end-to-end distance regulates the conformation statistics on the Kuhn scale. We then proceed to compute the time evolution of conformation statistics on the Kuhn scale at the start-up of simple flows.

In Section IV we extend the 2S2M model of entangled chains under flow to account for a self-consistent potential that approximates segmental interactions. We show how it modifies strand conformation statistics on the Kuhn scale, as the parameters controlling the persistent random walk are renormalized by multiplication with local statistical weights of *free* (unconnected) segment species. These can obtained as solutions of probabilistic transport equations in the self-consistent potential. Next, we invert the generalized Green-Kubo relation from the previous section, to determine how the second moment of strand's end-to-end distance is changed by the self-consistent potential. This allows us to devise a two-scale sime propagation scheme for the statistical weights of subchains of the entangled chain. The latter are used to calculate local volume fractions for each chemical type of Kuhn segments in entangled chains, thus determining the self-consistent potential. In Section V we conclude with a brief summary, and discuss the validity and interpretation of some of our approximations, the advantages and limitations of the proposed method, and possible future generalizations.

## II. IDEAL ENTANGLED STRANDS UNDER FLOW

### A. Evolution of Ideal Strand Conformations on Coarse Scale

In this section we limit our focus to dynamic conformations of an *ideal* (non-interacting) entangled strand under flow. Such a strand is approximated by a chain of exactly $N_{\mathrm{e}}^{\alpha}$ Kuhn segments of chemical type $\alpha$, connected by $N_{\mathrm{e}}^{\alpha}-1$ rigid, freely jointed links of length $a$ (the Kuhn length). We denote its end-to-end distance by $\mathbf{R}$, and $\langle \mathbf{RR} \rangle^\alpha$ is a symmetric second-rank tensor denoting the second moment of this quantity with respect to an arbitrary probability distribution over $\mathbf{R}$. At thermodynamic equilibrium, it assumes the isotropic form $\langle \mathbf{RR} \rangle_0^\alpha = \left(d_0^\alpha\right)^2 \boldsymbol{\delta}$, where $\langle \; \rangle_0^\alpha$ denotes averaging over an equilibrium ensemble of $\alpha$-type strands, $\boldsymbol{\delta}$ is the unit tensor and $d_0^\alpha = a\sqrt{N_\mathrm{e}^\alpha - 1}$. We define $\mathbf{A} = \langle \mathbf{RR} \rangle^\alpha / \left(d_0^\alpha\right)^2$ as the dimensionless form of $\langle \mathbf{RR} \rangle^\alpha$. At equilibrium, $\mathbf{A} = \frac{1}{3}\boldsymbol{\delta}$ and $\mathrm{tr}\,\mathbf{A} = 1$, so $A_{11} = A_{22} = A_{33} = \frac{1}{3}$ and $A_{12} = A_{13} = A_{23} = 0$.

In nonuniform flows, $\mathbf{A}$ deviates from its equilibrium form, and its time evolution combines affine deformation by the flow with entropy-driven relaxation. Here we adopt an approximate model for the time evolution of $\mathbf{A}$ in continuous, three-dimensional space, that was originally proposed by Marrucci and Ianniruberto [17]. According to their model,

$$\overset{\triangledown}{\mathbf{A}} = -\frac{f}{\tau}\left(\mathbf{A} - \tfrac{1}{3}\boldsymbol{\delta}\,\mathrm{tr}\,\mathbf{A}\right) - \frac{1}{\tau_\mathrm{R}}\left(f\,\mathrm{tr}\,\mathbf{A} - 1\right)\boldsymbol{\delta}, \tag{1}$$

where $\overset{\triangledown}{\mathbf{A}} = D\mathbf{A}/Dt - \mathbf{k}\cdot\mathbf{A} - \mathbf{A}\cdot\mathbf{k}^T$ is the upper-convected time derivative of $\mathbf{A}$, $D\mathbf{A}/Dt$ is its material derivative, $\mathbf{k} = \nabla\mathbf{u}$ is the velocity gradient tensor, and $\mathbf{k}^T$ is its transpose. In Eq. (1), the term $-\mathbf{k}\cdot\mathbf{A} - \mathbf{A}\cdot\mathbf{k}^T$ on the LHS describes affine deformation by the flow, combining stretching and compression along the principal axes of $\mathbf{A}$, as well as their rotation under flow. The affine deformation by the flow is counteracted by two entropy-driven relaxation terms on the RHS of Eq. (1). The first term drives relaxation of the orientation of the PP segment spanning the strand. The orientational relaxation time $\tau$ is defined by

$$\frac{1}{\tau} = \frac{2}{\tau_{\mathrm{d}}} + \left( \frac{1}{\tau_{\mathrm{R}}} - \frac{2}{\tau_{\mathrm{d}}} \right) \frac{\beta \left( f \, \mathrm{tr} \, \mathbf{A} - 1 \right)}{1 + \beta \left( f \, \mathrm{tr} \, \mathbf{A} - 1 \right)} . \tag{2}$$

The equation above couples the orientational relaxation time $\tau$ to the disengagement time $\tau_d$ of the Doi-Edwards reptation theory [15]. The factors 2 and $\left( f \, \mathrm{tr} \, \mathbf{A} - 1 \right)$ in Eq. (2) account for double reptation and convective constraint release (CCR), respectively. The second term on the RHS of Eq. (1) drives relaxation of strand's stretching and compression along the principal axes of $\mathbf{A}$. This occurs with the faster Rouse relaxation time $\tau_{\mathrm{R}} \ll \tau_{\mathrm{d}}$. Here

$$f = \frac{b - 1}{b - \mathrm{tr} \, \mathbf{A}} \tag{3}$$

where $b$ has the meaning of the square of the maximum stretch ratio, and $\beta$ is a numerical parameter of order unity measuring CCR effectiveness. Following Ref. [17], we use $\beta = 2$, and the square of the maximum stretch ratio is set here to $b = 100$.

The six independent components of $\mathbf{A}$, obtained by solving Eqs (1)-(3), were previously used to calculate transient and steady-state rheometrical response functions for a range of deformation rates in the case of simple elongational and shear flows [17], as well as for more complex flows [18]. It is evident that $\mathbf{A}(t)$ characterizes the evolution of conformation statistics of an ideal entangled strand on the coarse scale. We discuss below how it evolves at the start-up of simple elongational and shear flows, exploring a range of steady deformation rates. In the next section, we will show how $\mathbf{A}(t)$ regulates evolution of the statistics of ideal strand conformations on the Kuhn scale, which is much less obvious. The linear combination $\mathrm{tr} \, \mathbf{A} = A_{11} + A_{22} + A_{33}$ plays a special role in this regulation. Note that this combination can be expressed in the form $\mathrm{tr}\,\mathbf{A} = \lambda_1 + \lambda_2 + \lambda_3$ that is invariant under rotation, where $\lambda_1 \geq \lambda_2 \geq \lambda_3$ are the three eigenvalues of $\mathbf{A}$.

The eigenvector associated with the largest eigenvalue $\lambda_1$ is parallel to the flow-induced preferred orientation of the PP segment spanning the strand along the major axis of $\mathbf{A}$. In the case of the simple elongational flow considered in Figs. 2-4, the off-diagonal elements of $\nabla\mathbf{u}$ vanish, $\partial u_x / \partial x = \dot{\varepsilon}$, $\partial u_y / \partial y = \partial u_z / \partial z = -\dot{\varepsilon} / 2$, and the major axis of $\mathbf{A}$ coincides with the *x*-axis. On the other hand, for the simple shear flows considered in Figs. 2-4, the only component of $\nabla\mathbf{u}$ that does not vanish is $\partial u_x / \partial y = \dot{\gamma}$, and the major axis of $\mathbf{A}$ rotates in the *xy*-plane. The spatial extension of the strand along the major axis of $\mathbf{A}$ is $\sqrt{\lambda_1} d_0^\alpha$, where $d_0^\alpha = a\sqrt{N_\mathrm{e}^\alpha - 1}$, and is identified with the length of the PP segment spanning the strand. The eigenvectors associated with the smaller eigenvalues $\lambda_2$ and $\lambda_3$ set the orientations of the two minor axes of $\mathbf{A}$, which are orthogonal to the major axis and to each other. The mean spatial extensions of the strand along the two minor axes of $\mathbf{A}$ are $\sqrt{\lambda_2} d_0^\alpha$ and $\sqrt{\lambda_3} d_0^\alpha$.

Fig. 2 exhibits the evolution of $\mathrm{tr}\,\mathbf{A}$ with time for a range of different steady deformation rates at the start-up of simple elongational and shear flows. Fig. 3 shows the evolution of $\sqrt{\lambda_i}$ for the same flows as in Fig. 2. At $t = 0$, the fluid is at equilibrium state, where $\mathrm{tr}\,\mathbf{A} = 1$ and $\sqrt{\lambda_1} = \sqrt{\lambda_2} = \sqrt{\lambda_3} = \sqrt{\tfrac{1}{3}}$. As the time increases, $\mathrm{tr}\,\mathbf{A}$ and $\sqrt{\lambda_1}$ increase from their equilibrium values, stabilizing at higher steady-state values, which increase as the deformation rates are increased. This indicates that the flow stretches the length of the PP segment spanning the strand along the major axis of $\mathbf{A}$. Similarly, $\sqrt{\lambda_2}$ and $\sqrt{\lambda_3}$ decrease from their equilibrium values, stabilizing at reduced steady state values, which decrease as the deformation rates are increased. This indicates that the flow squeezes the spatial extension of the strand along the two minor axes. Note that $\sqrt{\lambda_2} = \sqrt{\lambda_3}$ for the elongational flows in Figs. 2-4, so we may think of $\sqrt{\lambda_2} d_0^{\alpha}$ as being proportional to the shrinking diameter of the confining tube. However, it is no longer appropriate to use the term "tube diameter" in connection with the shear flows in Figs. 2-4, since in this case $\sqrt{\lambda_2} > \sqrt{\lambda_3}$, so the cross-section of the tube diameter is no longer circular, but elliptic.

Comparing elongational and shear start-up flows in Figs. 2-3, it is interesting to note that, for elongational flows, the variation of $\mathrm{tr}\,\mathbf{A}$ and $\sqrt{\lambda_i}$ with time is monotonic, but it is non-monotonic for shear flows: in the latter case, $\mathrm{tr}\,\mathbf{A}$ and $\sqrt{\lambda_1}$ exhibit overshoots, while $\sqrt{\lambda_2}$ and $\sqrt{\lambda_3}$ exhibit undershoots, before stabilizing at their steady-state values. This indicates that some retraction of strand stretching and squeezing follows after they reach their extrema, before attaining the steady-state values.

Fig. 4 displays log-log plots of the values of $\sqrt{\lambda_i}$ as functions of the deformation rate at the steady state of simple flows. Note that in the limit of high deformation rates, $\sqrt{\lambda_1}$ approaches a constant value, set by the maximum extension ratio $\sqrt{b} = 10$, but $\sqrt{\lambda_2}$ and $\sqrt{\lambda_3}$ approach asymptotic straight lines of negative slope, indicating an asymptotic power law. Thus in the case of simple elongational flows, Fig. 4(a) indicates that $\sqrt{\lambda_3} = \sqrt{\lambda_2} \sim c_2 \left(\dot{\varepsilon}\tau_d\right)^{\alpha_2}$, where the slope is $\alpha_2 = -0.500$, up to three significant figures in our numerical computation. For simple shear flows, Fig. 4(b) shows that $\sqrt{\lambda}_2 \sim c_2 \left(\dot{\gamma}\tau_d\right)^{\alpha_2}$ and $\sqrt{\lambda}_3 \sim c_3 \left(\dot{\gamma}\tau_d\right)^{\alpha_3}$, where $c_3 < c_2$ but the two slopes have the equal value $\alpha_3 = \alpha_2 = -0.315$, up to three significant figures in our numerical computation.

**B. Conformations of Ideal Strands as a Persistent Random Walks on Kuhn Scale**

The coarse scale results presented above show significant stretching of the entangled strands along the major axis of $\mathbf{A}$ and shrinkage along the minor axes as deformation rates are increased. If random walks are used to sample conformation statistics of entangled strands on the Kuhn scale, this imposes correlations between the directions of successive steps, where each step corresponds to a rigid link between a pair of successive Kuhn segments along the strand. Such correlations are neglected by the Wiener random walks previously used to model unentangled chains in polymer fluids under flow. We construct below a persistent random walk model to sample conformation statistics of ideal entangled strands that accounts for correlations between the directions of successive steps, when the strands are stretched and oriented by the flow.

Consider a face-centered cubic (fcc) lattice, with the lattice constant being $a$, the Kuhn length of the polymer chain. Let $\mathbf{e}_1 = (1,0,0)$, $\mathbf{e}_2 = (0,1,0)$ and $\mathbf{e}_3 = (0,0,1)$ be a triad of unit vectors serving as the basis for the $(x,y,z)$ Cartesian system of coordinates. A site $\mathbf{r} = (x,y,z)$ belonging to the fcc lattice has 12 nearest-neighbors at $\mathbf{r} + \mathbf{a}_{\pm i}$, where $\mathbf{a}_{\pm i}$ is one of the following twelve unit vectors on the fcc lattice (see Fig. 5)

$$\begin{aligned}
&\mathbf{a}_1 = a\mathbf{e}_1, \quad \mathbf{a}_2 = a\left(\tfrac{1}{2}\mathbf{e}_1 + \tfrac{\sqrt{3}}{2}\mathbf{e}_2\right), \quad \mathbf{a}_3 = a\left(-\tfrac{1}{2}\mathbf{e}_1 + \tfrac{\sqrt{3}}{2}\mathbf{e}_2\right), \quad \mathbf{a}_4 = a\left(-\tfrac{1}{\sqrt{3}}\mathbf{e}_2 + \sqrt{\tfrac{2}{3}}\mathbf{e}_3\right), \\
&\mathbf{a}_5 = a\left(\tfrac{1}{2}\mathbf{e}_1 + \tfrac{1}{2\sqrt{3}}\mathbf{e}_2 + \sqrt{\tfrac{2}{3}}\mathbf{e}_3\right), \quad \mathbf{a}_6 = a\left(-\tfrac{1}{2}\mathbf{e}_1 + \tfrac{1}{2\sqrt{3}}\mathbf{e}_2 + \sqrt{\tfrac{2}{3}}\mathbf{e}_3\right), \quad \text{and} \quad \mathbf{a}_{-i} = -\mathbf{a}_i.
\end{aligned} \tag{4}$$

Let $P_i(\mathbf{r},s)$ be the probability that the random walker arrives at the site $\mathbf{r}$ at sime $s$, from the site $\mathbf{r}-\mathbf{a}_i$ that it occupied at sime $s-1$. A persistent random walk on the fcc lattice is defined by the following system of recursive equations for $i \in I_{12}$

$$P_i(\mathbf{r},s+1) = TP_i(\mathbf{r}-\mathbf{a}_i,s) + RP_{-i}(\mathbf{r}-\mathbf{a}_i,s) + 10LP_j(\mathbf{r}-\mathbf{a}_i,s), \tag{5}$$

where $I_{12} = \{\pm 1, \pm 2, \ldots, \pm 6\}$, generating $\{P_i(\mathbf{r},s+1)\}_{i\in I_{12}}$ from known probabilities $\{P_i(\mathbf{r},s)\}_{i\in I_{12}}$. A cubic lattice version [19,20] of Eq. (5) has been proposed as a persistent random walk model of such non-Fickian transport phenomena as mesoscopic diffusion, heat propagation, and anisotropic light scattering in crystalline lattices and disordered media. These equations describe a second order Markov chain [21], since they relate two successive sime steps: the preceding sime step between $s-1$ and $s$, and the next sime step between $s$ and $s+1$. Alternatively, they can be interpreted as the governing equations for a *multistate* first-order Markov chain [21], the 12 different states corresponding to the possible $s$-propagation directions on the fcc lattice . We identify $T$ as the transmission probability: the conditional probability for a walker that $s$-propagated along a direction $\mathbf{a}_i$ (from site $\mathbf{r}-2\mathbf{a}_i$ to site $\mathbf{r}-\mathbf{a}_i$) in the preceding sime step to continue $s$-propagating in this same direction (from $\mathbf{r}-\mathbf{a}_i$ to $\mathbf{r}$) in the next sime step. We identify $R$ as the reflection probability: the conditional probability for a walker that $s$-propagated in the opposite direction $-\mathbf{a}_i$ (from site $\mathbf{r}$ to site $\mathbf{r}-\mathbf{a}_i$ ) in the preceding sime step, will propagate in the direction of $\mathbf{a}_i$ (from $\mathbf{r}-\mathbf{a}_i$ to $\mathbf{r}$) in the next sime step. Similarly, $L$ is the lateral scattering probability: the conditional probability that a walker that propagated along any of the 10 directions $\mathbf{a}_j$, where $j \neq \pm i$ (from any of the 10 sites $\mathbf{r}-\mathbf{a}_i-\mathbf{a}_j$ to site $\mathbf{r}-\mathbf{a}_i$ ), in the preceding sime-step, will propagate in the direction of $\mathbf{a}_i$ in the next sime step. If the persistent random walk is invariant under lattice translations, then Eq. (5) assumes a simpler form

$$P_i\left(s+1\right)=\sum_{j\in I_{12}} M_{ij}P_j\left(s\right),\qquad i\in I_{12},\ s=1,\ldots,N_{\mathrm{e}}-1, \tag{6}$$

where

$$M_{ij}=\begin{cases} T & \text{if } i=j, \\ R & \text{if } i=-j, \\ L & \text{otherwise.}\end{cases} \tag{7}$$

In principle, Eq. (6) allows us to calculate the probability that the $s^{\text{th}}$ link in a subchain of the entanglement strand consisting of $s$ links is oriented along $\mathbf{a}_i$, provided that we know the values of $\left\{P_i\left(1\right)\right\}_{i\in I_{12}}$, the probability that the first link in this subchain is oriented along $\left\{\mathbf{a}_i\right\}_{i\in I_{12}}$. As we introduced our 2S2M model in Section I, we stressed that the notion of a localized, fixed entanglement position along the chain is only an approximation, since in reality the slip-link representing the entanglement is not localized, and fluctuates back and forth about a mean position. Our model reflects the properties of an average strand between entanglements, where such fluctuations are averaged out. Hence $P_i\left(1\right)$, the probability that the first link in any subchain of the entangled strand is oriented along $\mathbf{a}_i$, should be a sime-averaged quantity, and thus independent of its position along the strand. For this reason, we will use the notation $P_i\left(1\right)=\overline{p}_i$, where the bar indicates a sime average. We identify $\overline{p}_i$ with the probability that the PP segment of the tube spanned by the strand is oriented along $\mathbf{a}_i$, and note the symmetry $\bar{p}_{-i} = \bar{p}_i$. Substituting $\bar{p}_i$ instead of $P_i(1)$ in Eq. (6), we get

$$\bar{P}_i(s+1) = \sum_{j\in I_{12}} M_{ij}\bar{P}_j(s), \qquad s = 1,\ldots,N_\mathrm{e}-1, \tag{8}$$

where $i \in I_{12}$, and $\{\bar{P}_i(s)\}_{i\in I_{12}}$ are the probabilities that the $s^{\text{th}}$ link in a subchain of an entanglement strand is oriented along $\mathbf{a}_i$, averaged over the sime of the first link. Here $\bar{P}_i(s)$ satisfy the same symmetry $\bar{P}_{-i}(s) = \bar{P}_i(s)$. Eq. (8) is a system of coupled sime evolution equations for $\bar{P}_i(s)$, where the values of the sime-averaged orientational probabilities $\bar{p}_i$ specify the initial conditions. Note that for an ideal, non-interacting chain we have $\bar{p}_i = \frac{1}{12}$ for all $i \in I_{12}$. Eq. (8) can be written in a vector form $\bar{\mathbf{P}}(s+1) = \mathbf{M}\cdot\bar{\mathbf{P}}(s)$, where $\bar{\mathbf{P}}(1) = \bar{\mathbf{p}} = [\bar{p}_1, \bar{p}_{-1},\ldots,\bar{p}_6,\bar{p}_{-6}]^T$, and, for $s = 2,\ldots,N_\mathrm{e}-1$, $\bar{\mathbf{P}}(s)$ is the 12-component vector $[\bar{P}_1(s),\bar{P}_{-1}(s),\ldots,\bar{P}_6(s),\bar{P}_{-6}(s)]^T$. The elements $M_{ij}$ of the $12\times12$ "scattering" matrix $\mathbf{M}$ of transition "rates" are defined by Eq.(7). Since $\mathbf{M}$ is a *stochastic* matrix, the sum over the elements of each column must satisfy the normalization condition

$$T + R + 10L = 1. \tag{9}$$

As defined in Eq. (7), the matrix $\mathbf{M}$ has a simple *block-circulant* [22] form. It is also a *doubly stochastic* matrix, since the sum over the elements of each row satisfies Eq. (9) as well. Using the rules of matrix multiplication, the system of sime evolution equations in Eq. (8) can also be written as a single vector equation

$$\bar{\mathbf{P}}(s) = \mathbf{M}^{s-1}\cdot\bar{\mathbf{p}}, \qquad s = 2,\ldots,N_\mathrm{e}-1, \tag{10}$$

where the $12\times 12$ matrix $\mathbf{M}^{s-1}$ has the same block-circulant structure as $\mathbf{M}$. Its elements $\left(\mathbf{M}^{s-1}\right)_{ij}$ are the conditional probabilities that the last link in an $s$-link subchain of an entanglement strand is along $\mathbf{a}_i$, provided that its first link is along $\mathbf{a}_j$. They are easily calculated using matrix multiplication rules:

$$\left(\mathbf{M}^{s-1}\right)_{ij}=\begin{cases}\frac{5}{12}\left(T+R-2L\right)^{s-1}+\frac{1}{12}\left(T+R+10L\right)^{s-1}+\frac{1}{2}\left(T-R\right)^{s-1} & \text{if } i=j,\\ \frac{5}{12}\left(T+R-2L\right)^{s-1}+\frac{1}{12}\left(T+R+10L\right)^{s-1}-\frac{1}{2}\left(T-R\right)^{s-1} & \text{if } i=-j,\\ -\frac{1}{12}\left(T+R-2L\right)^{s-1}+\frac{1}{12}\left(T+R+10L\right)^{s-1} & \text{otherwise.}\end{cases} \tag{11}$$

## III. ADAPTIVE CONTROL OF THE PERSISTENT RANDOM WALK

### A. Generalized Green-Kubo Relation

We proceed now to show how the time evolution of $\mathbf{A}$ under flow regulates the parameters $\left\{\overline{p}_i\right\}_{i\in I_6}$ and $\left\{T,R,L\right\}$ that control segmental and conformation statistics of an ideal entangled strand on the Kuhn scale. Consider the conformations of a freely jointed subchain of a strand of $N_{\mathbf{e}}^{\alpha}$ Kuhn segments between successive entanglements, which are jointed by $N_{\mathbf{e}}^{\alpha}-1$ rigid links. In reality, the orientation of each link may be along any vector terminating on a unit sphere. In our model, we approximate the sampling of link orientation statistics by restricting the orientation of each link to one of the 12 unit vectors $\left\{\mathbf{a}_i\right\}_{i\in I_{12}}$ on the fcc lattice. The restricted chain conformations can be generated by a persistent random walker depositing $N_{\mathbf{e}}^{\alpha}-1$ rigid links connecting $N_{\mathbf{e}}^{\alpha}$ Kuhn segments along the nearest-neighbor bonds of the fcc lattice in $N_{\mathbf{e}}^{\alpha}-1$ successive sime steps. The $s$-displacement $\mathbf{d}\left(s\right)$ that the walker undergoes to deposit the $s^{\text{th}}$ link is oriented along one of the twelve vectors $\{\mathbf{a}_i\}_{i\in I_{12}}$. We assume a sime-averaged probability $\overline{p}_i$ that the first link in the subchain is oriented along $\mathbf{a}_i$, which corresponds to restricting the sampling of the orientations of the PP segments spanning this strand to one of the 12 vectors $\{\mathbf{a}_i\}_{i\in I_{12}}$ as well.

The persistent random walker requires $N_{\mathrm{e}}^{\alpha}-1$ sime steps to generate a conformation of the entire strand between successive entanglements, with a net $s$-displacement $\mathbf{R}=\sum_{s=1}^{N_{\mathrm{e}}^{\alpha}-1}\mathbf{d}(s)$. Hence the sime-averaged second moment of $\mathbf{R}$ is

$$\overline{\langle\mathbf{RR}\rangle^{\alpha}}=\overline{\left\langle\left[\sum_{s'=1}^{N_{\mathrm{e}}^{\alpha}-1}\mathbf{d}(s')\right]\left[\sum_{s''=1}^{N_{\mathrm{e}}^{\alpha}-1}\mathbf{d}(s'')\right]\right\rangle}=\sum_{s'=1}^{N_{\mathrm{e}}^{\alpha}-1}\left[\overline{\langle\mathbf{d}(s')\mathbf{d}(s')\rangle}+2\sum_{s''=2}^{N_{\mathrm{e}}^{\alpha}-s'}\overline{\langle\mathbf{d}(s')\mathbf{d}(s'')\rangle}\right],\quad(12)$$

where bars denote sime averaging of the pair autocorrelations with respect to the orientation of the first link in any subchain that belongs to the entangled strand.

The matrix elements $\left(\mathbf{M}^{s-1}\right)_{ij}$ are the conditional probabilities that the last link in an $s$-link subchain of an entanglement strand is directed along $\mathbf{a}_i$, provided that its first link is along $\mathbf{a}_j$. Hence $\left(\mathbf{M}^{s-1}\right)_{ij}\overline{p}_j$ is the joint probability that the $s^{\text{th}}$ link in a strand's subchain is oriented along $\mathbf{a}_i$, while its first link is oriented along $\mathbf{a}_j$, sime-averaged over the position of the first link along the strand. It is important to understand that, since the orientational probability of the first link is sime-averaged over its position along the strand, it must be invariant under sime translations of its position. This is why we regard it as a collective property of the strand, identified with the orientational probability of the PP segment that spans the strand. Hence the sime pair autocorrelations in Eq. (12), which are expectations of the dyadic products of the first and the terminal link vectors in a subchain of the strand, with respect to their joint probability distribution, are averaged over the sime of the first link as well. Therefore, these pair autocorrelations cannot depend on the sime of the first link, but rather depend only on the *difference* between the simes of the terminal link and the first link, resulting in the following relations:

$$\overline{\left\langle \mathbf{d}(s)\mathbf{d}(s)\right\rangle} = \overline{\left\langle \mathbf{d}(1)\mathbf{d}(1)\right\rangle} = \sum_{i\in I_{12}} \overline{p}_i \mathbf{a}_i \mathbf{a}_i, \tag{13}$$

and the sime-averaged pair autocorrelations at distinct simes satisfy

$$\overline{\left\langle \mathbf{d}(s')\mathbf{d}(s'')\right\rangle} = \overline{\left\langle \mathbf{d}(1)\mathbf{d}(s''-s')\right\rangle} = \sum_{i.\,j\in I_{12}} \left(M^{s''-s'}\right)_{ij} \overline{p}_j \mathbf{a}_i \mathbf{a}_j\ , \tag{14}$$

where $s''-s'=1,\ldots,N_{\mathrm{e}}^{\alpha}-2$. Using Eqs. (13) and (14), Eq. (12) reduces to the simplified form

$$\overline{\left\langle \mathbf{RR}\right\rangle^{\alpha}} = \left(N_{\mathbf{e}}^{\alpha}-1\right)\overline{\left\langle \mathbf{d}(1)\mathbf{d}(1)\right\rangle} + 2\sum_{s=2}^{N_{\mathrm{e}}^{\alpha}-1} \left(N_{\mathbf{e}}^{\alpha}-s\right)\overline{\left\langle \mathbf{d}(1)\mathbf{d}(s)\right\rangle}. \tag{15}$$

The equation above relates sime-averaged pair autocorrelations for microscopic $s$-displacements (by a single Kuhn length, representing a single link along the jointed chain representation of the strand) at equal and distinct simes of the persistent random walk on the fcc lattice, to the sime-averaged second moment $\overline{\left\langle \mathbf{RR}\right\rangle^{\alpha}}$ of its total $s$-displacement $\mathbf{R}$ after $N_{\mathrm{e}}^{\alpha}-1$ sime steps (corresponding to the end-to-end distance of the entangled strand). In Statistical Mechanics, such a relation is called the Green Kubo relation, after a similar relation between microscopic velocity pair autocorrelations and mean square displacements in fluids [23]. Using Eqs. (13)-(14) to evaluate autocorrelations in sime, Eq. (11) to evaluate $\left(M^{s''-s'}\right)_{ij}$, and the symmetry relations $\overline{p}_{-i} = \overline{p}_i$, we obtain

$$\overline{\langle \mathbf{RR} \rangle^{\alpha}} = \left(N_{\mathbf{e}}^{\alpha}-1\right)\sum_{i\in I_{12}} \bar{p}_i \mathbf{a}_i\mathbf{a}_i + 2\sum_{s=2}^{N_{\mathbf{e}}^{\alpha}-1}\left(N_{\mathbf{e}}^{\alpha}-s\right)\sum_{i,j\in I_{12}}\left(\mathbf{M}^{s-1}\right)_{ij}\bar{p}_j \mathbf{a}_i\mathbf{a}_j$$
$$= 2Z\left(N_{\mathbf{e}}^{\alpha}-1\right)\sum_{i\in I_6}\bar{p}_i\mathbf{a}_i\mathbf{a}_i \tag{16}$$

where $I_6 \equiv \{1, 2, \ldots, 6\}$,

$$Z = 1 + \frac{2\omega}{N_{\mathbf{e}}^{\alpha}-1}, \tag{17}$$

$$\omega = \sum_{s=2}^{N_{\mathbf{e}}^{\alpha}-1}\left(N_{\mathbf{e}}^{\alpha}-s\right)\Delta^{s-1}, \tag{18}$$

and $\Delta = T - R$. Recall that $\overline{\langle \mathbf{RR} \rangle^{\alpha}}$, the second moment of the strand's end-to-end distance in Eq. (16) is derived from a persistent random walk model that restricts the possible orientations of the successive links in an entanglement strand to the twelve unit vectors $\{\mathbf{a}_i\}_{i\in I_{12}}$ of the fcc lattice. We assume here that it is a good approximation for $\langle \mathbf{RR} \rangle^{\alpha}$ in a freely jointed chain model for conformations of entangled strands under flow, as obtained by solving Eqs. (1)-(3) (where link orientations are unrestricted). Then we can rewrite Eq. (16) as

$$\mathbf{A} = Za^{-2}\langle \mathbf{a}_i\mathbf{a}_i \rangle_{i\in I_{12}} = Z\boldsymbol{\sigma}, \tag{19}$$

where $\boldsymbol{\sigma} = a^{-2}\langle \mathbf{a}_i\mathbf{a}_i \rangle_{i\in I_{12}}$ is the dimensionless second moment of the discrete distribution of the orientations of the PP segments, restricted to the twelve directions $\{\mathbf{a}_i\}_{i\in I_{12}}$ on the fcc lattice, and $Z$ is an overall scaling factor.

**B. Evolution of Orientational Probabilities and Orientational Entropy**

Using Eqs(4), the second-rank tensor $\boldsymbol{\sigma}$ assumes the form

$$
\begin{aligned}
\boldsymbol{\sigma} = a^{-2}\sum_{i\in I_{12}} \overline{p}_i \mathbf{a}_i\mathbf{a}_i &= \tfrac{1}{2}\left(4\overline{p}_1 + \overline{p}_2 + \overline{p}_3 + \overline{p}_5 + \overline{p}_6\right)\mathbf{e}_1\mathbf{e}_1 \\
&+ \tfrac{1}{6}\left(9\overline{p}_2 + 9\overline{p}_3 + 4\overline{p}_4 + \overline{p}_5 + \overline{p}_6\right)\mathbf{e}_2\mathbf{e}_2 + \tfrac{4}{3}\left(\overline{p}_4 + \overline{p}_5 + \overline{p}_6\right)\mathbf{e}_3\mathbf{e}_3 \\
&+ \tfrac{1}{2\sqrt{3}}\left(3\overline{p}_2 - 3\overline{p}_3 + \overline{p}_5 - \overline{p}_6\right)\left(\mathbf{e}_1\mathbf{e}_2 + \mathbf{e}_2\mathbf{e}_1\right) \\
&+ \sqrt{\tfrac{2}{3}}\left(\overline{p}_5 - \overline{p}_6\right)\left(\mathbf{e}_1\mathbf{e}_3 + \mathbf{e}_3\mathbf{e}_1\right) + \tfrac{\sqrt{2}}{3}\left(2\overline{p}_4 - \overline{p}_5 - \overline{p}_6\right)\left(\mathbf{e}_2\mathbf{e}_3 + \mathbf{e}_3\mathbf{e}_2\right).
\end{aligned} \tag{20}
$$

Substituting Eq. (21) into Eq.(19), we get the following system of six linear equations for $\{\overline{p}_i\}_{i\in I_6}$

$$
\begin{aligned}
&\tfrac{1}{2} Z\left(4\overline{p}_1 + \overline{p}_2 + \overline{p}_3 + \overline{p}_5 + \overline{p}_6\right) = A_{11}, \\
&\tfrac{1}{6} Z\left(9\overline{p}_2 + 9\overline{p}_3 + 4\overline{p}_4 + \overline{p}_5 + \overline{p}_6\right) = A_{22}, \\
&\tfrac{4}{3} Z\left(\overline{p}_4 + \overline{p}_5 + \overline{p}_6\right) = A_{33}, \\
&\tfrac{1}{2\sqrt{3}} Z\left(3\overline{p}_2 - 3\overline{p}_3 + \overline{p}_5 - \overline{p}_6\right) = A_{12}, \\
&\tfrac{\sqrt{2}}{3} Z\left(-2\overline{p}_4 + \overline{p}_5 + \overline{p}_6\right) = A_{23}, \\
&\sqrt{\tfrac{3}{2}} Z\left(\overline{p}_5 - \overline{p}_6\right) = A_{13}.
\end{aligned} \tag{22}
$$

The solution of the system of equation above is a function of $\mathbf{A}$ and $Z$ :

$$
\begin{aligned}
\overline{p}_1 &= \left(\tfrac{1}{2}A_{11} - \tfrac{1}{6}A_{22} - \tfrac{1}{12}A_{33} - \tfrac{1}{3\sqrt{2}}A_{23}\right)/Z, \\
\overline{p}_2 &= \left(\tfrac{1}{3}A_{22} - \tfrac{1}{12}A_{33} + \tfrac{1}{\sqrt{3}}A_{12} + \tfrac{1}{6\sqrt{2}}A_{23} - \tfrac{1}{2\sqrt{6}}A_{13}\right)/Z, \\
\overline{p}_3 &= \left(\tfrac{1}{3}A_{22} - \tfrac{1}{12}A_{33} - \tfrac{1}{\sqrt{3}}A_{12} + \tfrac{1}{6\sqrt{2}}A_{23} + \tfrac{1}{2\sqrt{6}}A_{13}\right)/Z, \\
\overline{p}_4 &= \left(\tfrac{1}{4}A_{33} - \tfrac{1}{\sqrt{2}}A_{23}\right)/Z, \\
\overline{p}_5 &= \left(\tfrac{1}{4}A_{33} + \tfrac{1}{2\sqrt{2}}A_{23} + \tfrac{\sqrt{6}}{4}A_{13}\right)/Z, \\
\overline{p}_6 &= \left(\tfrac{1}{4}A_{33} + \tfrac{1}{2\sqrt{2}}A_{23} - \tfrac{\sqrt{6}}{4}A_{13}\right)/Z.
\end{aligned} \tag{23}
$$

where the six independent components of the second-rank tensor $\mathbf{A}(t)$ are the solutions of the differential evolution equations (Eqs. (1)-(3)) at time $t$. The parameter Z is determined from the normalization condition

$$
\sum_{i\in I_{12}} \overline{p}_i = 2\sum_{i\in I_6} \overline{p}_i = 1, \tag{24}
$$

for sime-averaged orientational probabilities (where we used $\bar{p}_{-i} = \bar{p}_i$), giving $Z = \mathrm{tr}\,\mathbf{A}$. Substituting the last equation into Eq.(23), we see that, at any time $t$, the orientational probabilities $\{\bar{p}_i(t)\}_{i\in I_6}$ in our Kuhn scale lattice model for the entangled strands are completely determined by $\mathbf{A}(t)$. It is interesting to consider the regime where $Z = 1$, which is valid at the steady states of simple elongational and shear flows when $\dot{\varepsilon}\tau_{\mathrm{d}}, \dot{\gamma}\tau_{\mathrm{d}} \ll 1$, or at early times after start-up (see Figs. 2 – 4). In this case the system of equations for $\bar{p}_i$ in Eq. (22), and its solution $\bar{p}_i(\mathbf{A})$ in Eq. (23) become identical to Eqs. (24) and (27) of Ref. [1] that determine the dependence of the anisotropic stepping probabilities in a Kuhn-scale Wiener random walk sampling the conformations of unentangled ideal chains under flow, on the dimensionless second moment of their end-to-end distance (after adjusting for a different indexing scheme for $\{\mathbf{a}_i\}_{i\in I_{12}}$ in Ref. [1]). This means that in the regime $Z = 1$ of our 2S2M model for the ideal entangled chain under flow, the PP segments correspond to steps of an anisotropic Wiener random walk sampling PP conformations on the PP segment scale, where there are no correlations between the directions of successive steps (as in the independent alignment approximation of the Doi-Edwards model [15]).

Fig. 6 shows the time evolution of the orientational probabilities $\{\bar{p}_i\}_{i\in I_{12}}$ of ideal entangled strands at the start-up of simple flows with very high, but finite, deformation rates. Note that in the equilibrium limit $t \to 0^+$, all twelve orientational probabilities are identical, $\bar{p}_i = \frac{1}{12}$, as expected for an isotropic distribution. Fig. 6(a) shows $\bar{p}_i(t)$ for elongational flow with dimensionless elongation rate $\dot{\varepsilon}\tau_d = 500$, approaching the steady-state values $\bar{p}_{\pm 1} \to \frac{1}{2}$ and $\bar{p}_{\pm i} = 0$ for $i \neq \pm 1$. Fig. 6(b) displays $\bar{p}_i(t)$ in the case of shear flow with dimensionless shear rate $\dot{\gamma}\tau_d = 1000$, approaching the steady state $\bar{p}_{\pm 1} \to 0.491$, $\bar{p}_{\pm 2} \to 0.036$, $\bar{p}_{\pm 4} = \bar{p}_{\pm 5} = \bar{p}_{\pm 6} = 0.002$, and $\bar{p}_{\pm 3} \to -0.032 < 0$. Non-monotonic time dependence is seen in $\bar{p}_2(t)$ and $\bar{p}_3(t)$ that can be attributed to non-monotonic time dependence of $\sqrt{\lambda_i(t)}$ and $\theta(t)$ in Figs 3(b) and 3(c).

Let us now consider the high deformation rate limit of the second-rank tensor $\mathbf{A}$ corresponding to the dimensionless second moment of the end-to-end distance of the off-lattice strands, obtained from the steady state solutions of Eqs. (1)-(3), in the case of the two simple flows considered in Figs. 2-4 and Fig. 6. In this limit, $Z = \mathrm{tr}\,\mathbf{A} \to b$ and $\mathbf{A} \to b\mathbf{e}_1\mathbf{e}_1$ at the steady state of both elongational and shear flows, since in this regime $\mathbf{A}$ is diagonal, the eigenvector associated with its largest eigenvalue $\lambda_1$ is aligned along $\mathbf{e}_1$, $\lambda_1 \to b$ and the remaining eigenvalues $\lambda_2, \lambda_3 \to 0$ (here we assumed $b = 100$). However, $\mathbf{A}(t)$ remains diagonal at all times during elongational start-up flows, while the eigenvector associate with $\lambda_1$ rotates from being aligned along the unit vector $(\mathbf{e}_1 + \mathbf{e}_2)/\sqrt{2}$ that is oriented at $45^0$ angle between $\mathbf{e}_1$ and $\mathbf{e}_2$ in the $xy$ plane, toward a direction aligned with $\mathbf{e}_1$, as the steady state is approached. At intermediate times, when this rotation takes place, $\bar{p}_{\pm 3}$ assume negative values. Looking at the plot in Fig. 3(c) corresponding to the dimensionless shear rate $\dot{\gamma}\tau_\mathrm{d} = 1000$, we see that the eigenvector associated with $\lambda_1$ gets close to, but not fully aligned with the direction of $\mathbf{e}_1$ at this high, but finite shear rate. Comparing Fig. 6(b) with Fig. 3(c), we see that, for $\dot{\gamma}\tau_\mathrm{d} = 1000$, $\bar{p}_{\pm 3}$ attain a minimum negative value around the time when the mismatch between the direction of the eigenvector associated with $\lambda_1$, and the twelve fcc lattice vectors $\{\mathbf{a}_i\}_{i\in I_{12}}$, is maximal (i.e. when $\theta(t)\approx 30^0$), and subsequently increases back toward zero, though stabilizing at a finite negative steady-state value.

What forces $\{\overline{p}_i\}_{i\in I_{12}}$ to assume values outside the interval $[0,1]$ in the case of shear flow at high shear rates? For such flows, if $\{\overline{p}_i\}_{i\in I_{12}}$ are restricted to the interval $[0,1]$ there is a mismatch between the eigenvectors of $\boldsymbol{\sigma}=a^{-2}\langle\mathbf{a}_i\mathbf{a}_i\rangle_{i\in I_{12}}$ (where the dimensionless second moment is over orientations that are restricted to the twelve directions $\{\mathbf{a}_i\}_{i\in I_{12}}$ on the fcc lattice), and the eigenvectors of $\mathbf{A}=d_0^{-2}\langle\mathbf{RR}\rangle$ ( where the second moment is obtained by solving Eqs. (1)-(3), which are free of lattice restrictions). When such mismatch occurs with the restriction $\overline{p}_i\in[0,1]$ for all $i$, the Green-Kubo relation is *frustrated* by the lattice approximation. There is no such lattice frustration of the Green-Kubo relation in the case of elongational flows considered here, since the largest eigenvector is always along the lattice vector $\mathbf{a}_1$. In principle, we could reduce the lattice frustration for shear flows, as well as for more complex flows, by increasing the number of allowed directions for PP segments of entangled strands in the lattice model. However, this would come at a great computational cost that we are not willing to pay, as computational efficiency is one of our main goals.

According to the classical definition of a probability measure, its values must be confined to the interval $[0,1]$, so that it can be expressed as a limit of relative frequency ratios for observable events. Hence, we cannot use the term “probabilities” for $\{\overline{p}_i\}_{i\in I_{12}}$ when they assume real values outside the interval $[0,1]$, as is the case in our lattice model for simple shear flows at high enough shear rates. Following precedents in the published literature, we will refer to $\{\overline{p}_i\}_{i \in I_{12}}$ as *extended* (signed) probabilities [24] when they assume real values outside the interval $[0,1]$. Ordinary stochastic processes are defined only for classical probability measures that are restricted to the interval $[0,1]$. However when time-dependent processes satisfy relations formally identical to those satisfied by ordinary stochastic process, except that they relate signed (extended) probability measures, they are called stochastic *pseudo*-processes, or Markov pseudo-processes, if they also satisfy the Markov property. Such signed probability measures appear in Markov pseudo-processes associated with solutions of certain differential equations with higher than second order spatial derivative terms, that have been studied in probability theory since the pioneering work of Krylov [25]. These equations can be either of the heat/diffusion (parabolic) type [26], as well as of the telegrapher's (hyperbolic) type [27]. Thus when $\overline{p}_{\pm 3}$ become negative in our lattice model sampling the conformation statistics on the Kuhn scale (as is the case in Fig. 6(b)), the governing persistent random walk equations should be regarded as a multistate Markov pseudo-process. Extended (signed) probability measures and stochastic pseudo-processes have also been useful for analyzing other physical phenomena, where they arise due to physical or boundary constraints. We postpone a discussion of the similarities between these phenomena and the extended orientational probabilities regulating the associated Markov pseudo process in our lattice model to Section V. Following these examples, we will track and use the extended orientational probabilities for the entangled strands, and the associated Markov pseudo-process, as book-keeping devices that greatly facilitate intermediate calculations in our approximate model.

We need an entropic measure for the degree of orientational ordering of the ideal entangled strands under flow, as it is used in the probabilistic transport equations for the statistical weight of the free segments and segmental momentum, in a similar fashion to that of DSCF of unentangled polymer fluids [1]. If $0 \le \bar{p}_i \le 1$ for all $i$, the Shannon entropy

$$S_{\bar{\mathbf{p}}}(t) = -\sum_{i \in I_{12}} p_i(t) \log p_i(t) = -2\sum_{i \in I_6} p_i(t) \log p_i(t)_i \tag{25}$$

of the orientational probability distribution $\{\bar{p}_i\}_{i \in I_6}$ is real, and can be used to calculate the effect of the orientational ordering of the entangled strands on the segmental transport equations, which are similar to those derived in Ref. [1] for unentangled chains. However, $S_{\bar{\mathbf{p}}}(t)$ becomes complex if $\bar{p}_i$ is becomes negative for some $i$, as is the case approaching the steady state in Fig. 6(b). Hence we need an alternative entropic measure for the degree of orientational ordering that remains real and positive in such cases, and is a good approximation to the Shannon entropy when all $\bar{p}_i$ remain non-negative. For this purpose we consider two families of alternative entropy measures that were proposed in the literature. Each such family of entropy measures is parameterized by a continuous parameter $q$. The first family of generalized entropy measures is called the Renyi entropy of order $q$ [28], defined for the extended orientational probability $\bar{p}_i$ as

$$S^R_{q\bar{\mathbf{p}}} = \frac{1}{1-q} \ln\left( \sum_{i \in I_{12}} \bar{p}_i^q \right) = \frac{1}{1-q} \ln\left( 2\sum_{i \in I_6} \bar{p}_i^q \right) \tag{26}$$

The second family of generalized entropy measures is called the Tsallis entropy of order $q$ [29], defined for the extended orientational probability $\bar{p}_i$ as

$$S_{q\bar{\mathbf{p}}}^{T}=\frac{1}{1-q}\left(\sum_{i\in I_{12}}\bar{p}_i^{q}-1\right)=\frac{1}{1-q}\left(2\sum_{i\in I_6}\bar{p}_i^{q}-1\right) \tag{27}$$

Note that the Tsallis entropy and the Renyi entropy of the same order $q$ are monotonically related [30],

$$S_{q\bar{\mathbf{p}}}^{R}=\frac{\ln\left[1+(1-q)S_{q\bar{\mathbf{p}}}^{T}\right]}{1-q} \quad \text{or} \quad S_{q\bar{\mathbf{p}}}^{T}=\frac{e^{(1-q)S_{q\bar{\mathbf{p}}}^{R}}-1}{1-q}. \tag{28}$$

Hence $S_q^R$ and $S_{q\bar{\mathbf{p}}}^T$ will attain their maximal values at the same values of $\bar{\mathbf{p}}$. Both $S_q^R$ and $S_{q\bar{\mathbf{p}}}^T$ are real and positive for even orders $q>1$ (even if $\bar{p}_i$ is negative for some $i$), and they both reduce to the Shannon entropy in the limit $q\to 1$. The choice $q=2$ is the even order that is closest to the Shannon limit. The second-order Renyi entropy for the orientational probability distribution $\bar{\mathbf{p}}$ assumes the form

$$S_{2\bar{\mathbf{p}}}^{R}=-\ln\left(\sum_{i\in I_{12}}\bar{p}_i^{2}\right)=-\ln 2\left(\sum_{i\in I_6}\bar{p}_i^{2}\right), \tag{29}$$

The second-order Tsallis entropy for the orientational probability distribution $\bar{\mathbf{p}}$ has the form

$$S_2^T=1-\sum_{i\in I_{12}}\bar{p}_i^2=1-2\sum_{i\in I_6}\bar{p}_i^2\,, \tag{30}$$

and sometimes is referred to as the linearized entropy.

At thermodynamic equilibrium, $\bar{p}_{\pm i}=\frac{1}{12}$ for all $i\in I_6$. Thus the equilibrium limit of the second-order Renyi entropy is $\ln 12$, which is identical to the equilibrium limit of the Shannon entropy. In contrast, the equilibrium limit of the second-order Tsallis entropy is $\frac{11}{12}$. Fig. 7 compares the Shannon entropy and the second-order Renyi and Tsallis entropies of the extended orientational probability distributions $\bar{\mathbf{p}}(t)$ for the simple flows shown in Fig. 6. At the start-up of elongational flow (Fig. 7(a)), the Shannon entropy remains real and positive, while decreasing from the equilibrium value and smoothly approaching a greatly reduced value at the steady state, indicating strong orientational ordering of the strands in this process. The second-order Renyi entropy shows very little quantitative difference from the Shannon entropy in this case. The second-order Tsallis entropy exhibits a very similar qualitative shape to both Shannon and the second-order Renyi entropies, but on a different scale. Fig. 7(b) shows the time evolution of the same three entropic measures at the start-up of a simple shear flow at a high shear rate. Here as well, the Shannon and the second-order Renyi entropies initially exhibit little quantitative difference. However, the Shannon entropy becomes complex after the time at which $\overline{\mathbf{p}}_{\pm 3}$ turn negative in Fig. 6(b); hence Fig. 7(b) does not exhibit the Shannon entropy after this time. However, the second-order Renyi and Tsallis entropies both remain positive after this time and smoothly approach a steady state at a greatly reduced value, indicating strong orientational ordering of the strands in this process. Thus replacing the Shannon entropy by the second-order versions of either Renyi or Tsallis entropies avoids computational difficulties associated with complex-valued entropy measures. In comparison to the second-order Tsallis entropy, the second-order Renyi entropy seems to be a better approximation to the Shannon entropy when the latter is real-valued. Similarly to the Shannon entropy, but unlike the Tsallis entropy, the Renyi entropy is also an *additive* entropy measure [31], which may be a required property if we wish to make the connection to thermodynamic entropy. These considerations lead us to conclude that among these three entropic measures, the second-order Renyi entropy is the best choice for the entropic measure of the orientational extended probability distribution $\overline{\mathbf{p}}(t)$.

In Fig. 8 we show the time evolution of the second-order Renyi entropy for the orientational extended probability distribution $\bar{\mathbf{p}}(t)$ at the start-up of simple flows as the deformation rate is increased. Figs. 8(a) and 8(b) correspond to simple elongational and shear flows, respectively. In both figures, the second-order Renyi entropies decrease from the equilibrium value of $\ln 12$ and stabilize at a lower steady-state value that decreases with increasing deformation rates. However, $S_{2\bar{\mathbf{p}}}^{R}(t)$ in Fig. 8b exhibit some non-monotonic time dependence that can be traced to non-monotonicity of $\sqrt{\lambda_i(t)}$ in Fig. 3(b).

### C. Evolution of Scattering Probabilities and of the Scattering Entropy Rate

In the previous section we used the orientational (extended) probability distribution $\bar{\mathbf{p}}(t)$ and its associated entropy measure $S_{2\bar{\mathbf{p}}}^{R}(t)$ to quantify the orientation of ideal strands under flow, approximated as persistent random walks on the fcc lattice. We used the Green-Kubo relation derived in subsection III.A to calculate $\bar{\mathbf{p}}(t)$ and $S_{2\bar{\mathbf{p}}}^{R}(t)$ from the values of $\mathbf{A}(t)$ obtained by solving the continuum differential evolution equation (Eqs. (1)-(3)). Here we show how to use the same values of $\mathbf{A}(t)$ and the Maximum Entropy Principle to determine the time evolution of the transmission ($T$), reflection ($R$) and lateral scattering ($L$) components of the scattering matrix $\mathbf{M}$ and its associated entropy rate $S_{\mathbf{M}}$ [32]. The latter is needed to quantify the stretching of ideal strands under flow, approximated as persistent random walks on the fcc lattice.

Let us define the auxiliary function

$$F(\Delta) = \sum_{s=2}^{N_e - 1} \Delta^s = \frac{\Delta^2 - \Delta^{N_e^\alpha}}{1-\Delta} \approx \frac{\Delta^2}{1-\Delta}, \tag{31}$$

where $\Delta = T - R < 1$, and the term $\Delta^{N_\mathrm{e}^\alpha}$ can be neglected since $\Delta^{N_\mathrm{e}^\alpha} \ll \Delta^2$ when $N_\mathrm{e}^\alpha >> 2$ (here we assume $N_\mathrm{e}^\alpha = 50$). From Eqs. (18) and (31), we get

$$\omega = \frac{N_\mathrm{e}^\alpha F}{\Delta} - \frac{\partial F}{\partial \Delta} \approx \frac{N_\mathrm{e}^\alpha \left(1-\Delta^2\right) - 2\Delta - \Delta^2}{\left(1-\Delta\right)^2}. \tag{32}$$

The two solutions of this quadratic equation for $\Delta$ are

$$\Delta = \frac{N_\mathrm{e}^\alpha - 2 + 2\omega \pm \left[\left(N_\mathrm{e}^\alpha - 2 + 2\omega\right)^2 - 4\omega\left(N_\mathrm{e}^\alpha + 1 + \omega\right)\right]^{1/2}}{2\left(N_\mathrm{e}^\alpha + 1 + \omega\right)}, \tag{33}$$

where $\omega = \left(Z-1\right)/2$, and $Z = \mathrm{tr}\,\mathbf{A}$. At thermodynamic equilibrium, $Z = 1$, $\omega = 0$, and $\Delta = 0$, so the physical solution corresponds to choosing the negative sign in front of the square root above.

We know that the scattering probabilities $T$, $R$ and $L$ must satisfy two constraints. The first constraint is

$$T - R = \Delta, \tag{34}$$

where $\Delta$ is a function of $N_\mathrm{e}^\alpha$ and $\mathbf{A}$ that is determined by the physical branch of Eq. (33). The second constraint is provided by the normalization constraint on the scattering probabilities, Eq. (9). As we are not aware of any other independent constraints on $T$, $R$ and $L$, we rely on the Maximum Entropy Principle to determine an additional equation closing the system of three independent equations that can be solved to determine the three scattering probabilities. For this purpose note that $S_\mathbf{M}$, the entropy rate of the persistent random walk on the fcc lattice [32]

$$S_\mathbf{M} = -\sum_{i \in I_{12}} \mu_i \sum_{j \in I_{12}} M_{ij} \ln M_{ij} \tag{35}$$

where $\mu_i = \frac{1}{12}$ are the components of the stationary probability distribution corresponding to the normalized eigenvector $\boldsymbol{\mu}$ of $\mathbf{M}$ associated with the largest eigenvalue $\lambda = 1$. Hence in this case the entropy rate $S_{\mathbf{M}}$ [32] is identical to the Shannon entropy of the scattering probability distribution $\{T, R, L, \ldots, L\}$:

$$S_{\mathbf{M}} = -T \ln T - R \ln R - 10 L \ln L \,. \tag{36}$$

According to the Maximum Entropy Principle, we can determine $T$, $R$, and $L$ by maximizing $S_{\mathbf{M}}$ subject to the constraints in Eqs. (9) and (34). The latter allow us to express $T$ and $R$ as

$$T = \left(1 - 10L + \Delta\right)/2, \qquad R = \left(1 - 10L - \Delta\right)/2 \,. \tag{37}$$

Substituting Eq. (37) into Eq. (36), we notice the following necessary condition for $S_{\mathbf{M}}$ to be at a maximum:

$$\ln\left(\frac{1-10L+\Delta}{2}\right) + \ln\left(\frac{1-10L-\Delta}{2}\right) = 2 \ln L \,, \tag{38}$$

which can be reduced to a quadratic equation for $L$,

$$\left(1-10L+\Delta\right)\left(1-10L-\Delta\right) = 4L^2 \,. \tag{39}$$

This quadratic equation has two solutions,

$$L = \frac{5 \pm \sqrt{1+24\Delta^2}}{48} \,. \tag{40}$$

At thermodynamic equilibrium $\Delta = 0$ and $T = R = L = \frac{1}{12}$, so the physical solution must have a negative sign in front of the square root in Eq. (40).Substituting back into Eqs. (9) and (34), we get $T(\Delta)$ and $R(\Delta)$. Since $Z = \operatorname{tr} \mathbf{A}$, Eqs. (17) and (33) determine $\Delta(\operatorname{tr} \mathbf{A})$, allowing us to compute $T$, $R$ and $L$ as a function of $\operatorname{tr} \mathbf{A}$, as shown in Fig. 9.

Since $\mathbf{A}(t)$ is obtained as the solution of the continuum differential evolution equation (Eqs. (1)-(3)), this allows us to determine the time evolution of $T$, $R$ and $L$ at the start-up of simple flows, as shown in Fig. 10. Note that, at thermodynamic equilibrium, $\mathrm{tr}\,\mathbf{A} = 1$, and $\omega = \Delta = 0$, giving $T = R = L = \frac{1}{12}$ at $t = 0$. As the strand is stretched between successive entanglement points at the start-up of simple flows, $Z = \mathrm{tr}\,\mathbf{A}$ increases from 1, and both $\omega$ and $\Delta = T - R$ increase from zero. It is interesting that increase in $T$ (decrease in $R$ and $L$) is a monotonic function of time for all the elongation rates shown in Fig. 10(a), while it becomes a non-monotonic function of time (exhibits overshoots) at the higher shear rates shown in Fig. 10(b). This mirrors the non-monotonic behavior exhibited by $\sqrt{\lambda_1}$ and $\sqrt{\lambda_3}$ as a function of time in the case of simple shear flows (Fig. 3(b)), but not in the case of simple elongational flows.

Fig. 11 shows the time evolution of the scattering entropy rates $S_{\mathbf{M}}$ at the start-up of the same simple flows as in Fig. 10. Note that the scattering entropy rate attains its highest value $S_{\mathbf{M}} = \ln 12$ at thermodynamic equilibrium ($t = 0$), and stabilizes at a reduced steady-state values indicating ordering due to stretching. Larger deformation rates have smaller steady-state scattering entropy rates. However, $S_{\mathbf{M}}(t)$ is a monotonic function of time in the case of simple elongational flows (Fig. 11(a)), while in the case of simple shear flows (Fig. 11(b)) it is a non-monotonic function (exhibits overshoots).

## IV. INTERACTING ENTANGLED CHAINS UNDER FLOW

### A. Entangled Strands as Persistent Random Walks in a Self-Consistent Potential

So far we focused on how conformation statistics of entangled strands in ideal (noninteracting) entangled chains are affected by nonuniform flow. First, we solved a differential evolution equation to determine the dimensionless second moment of strand's end-to-end distance $\mathbf{A}$ as a function of the local velocity gradient $\nabla\mathbf{u}$. Then the orientational probabilities $\{\overline{p}_i\}_{i\in I_6}$ and the scattering probabilities $T$, $R$ and $L$ controlling the persistent random walk that samples the conformation statistics of ideal strands on the Kuhn-scale fcc lattice were determined from the known components of $\mathbf{A}$, using the generalized Green-Kubo relation and the Maximum Entropy Principle.

Here we are concerned with the additional effect on conformation statistics of Flory-type segmental interactions with chemically distinct adjacent segments and walls, approximated by a self-consistent potential varying on the Kuhn scale at the interfaces. Interaction with the self-consistent potential affects the statistical weight $\tilde{n}^{\alpha}(\mathbf{r})$ for placing a *free* Kuhn segment of type $\alpha$ at the destination site $\mathbf{r}$, as the persistent random walk generates strand conformations on the fcc lattice. Note that $\tilde{n}^{\alpha}(\mathbf{r})>1$ $\left(\tilde{n}^{\alpha}(\mathbf{r})<1\right)$ if the interaction of the free segment with the self-consistent field at $\mathbf{r}$ is attractive (repulsive), and $\tilde{n}^{\alpha}(\mathbf{r})=1$ if the self-consistent field at $\mathbf{r}$ is zero. It is possible to compute $n^{\alpha}(\mathbf{r})$ by solving probabilistic transport equations for free segment probabilities in a self-consistent potential, as was done in the DSCF theory for unentangled polymer fluids [1], but modified for entangled dynamics.

Accounting for the interaction with the self consistent field at the destination site $\mathbf{r}$, the statistical weight $\tilde{\overline{P}}_i^{\alpha}(\mathbf{r},s+1)$ for placing a Kuhn segment to terminate an $\alpha$-type jointed subchain of length $s+1$ at site $\mathbf{r}$ (arriving from an $\alpha$-type jointed subchain of length $s$ at $\mathbf{r}-\mathbf{a}_i$), is defined recursively, similarly to Eq. (6),

$$\tilde{\bar{P}}_i^{\alpha}\left(\mathbf{r}, s+1\right) = \sum_{j \in I_{12}} \tilde{M}_{ij}^{\alpha}\left(\mathbf{r}-\mathbf{a}_i\right) \tilde{\bar{P}}_j^{\alpha}\left(\mathbf{r}-\mathbf{a}_i, s\right) \tag{41}$$

where $i \in I_{12}$ and $s = 1, \dots, N_{\mathrm{e}}^{\alpha} - 1$. Here

$$\tilde{M}_{ij}^{\alpha}\left(\mathbf{r}-\mathbf{a}_i\right) = \tilde{n}^{\alpha}\left(\mathbf{r}\right) M_{ij}\left(\mathbf{r}-\mathbf{a}_i\right) \tag{42}$$

are the *renormalized* scattering probabilities that are obtained from the ideal-strand scattering probabilities $M_{ij}\left(\mathbf{r}-\mathbf{a}_i\right)$ by multiplication with the statistical weight $\tilde{n}^{\alpha}\left(\mathbf{r}\right)$ at the destination site. The sime-averaged probabilities for the orientation of the PP segment spanning the entangled strands are also renormalized in the presence of the self consistent potential:

$$\tilde{\bar{p}}_i^{\alpha}\left(\mathbf{r}\right) = \tilde{\bar{P}}_i^{\alpha}\left(\mathbf{r}, 1\right) = \tilde{n}^{\alpha}\left(\mathbf{r}\right) \bar{p}_i\left(\mathbf{r}\right). \tag{43}$$

The total statistical weight for finding a terminal segment of a strand subchain of $s$ segments at site $\mathbf{r}$ (regardless of arrival direction) is

$$\tilde{\bar{P}}^{\alpha}\left(\mathbf{r}, s\right) = \sum_{i \in I_{12}} \tilde{\bar{P}}_i^{\alpha}\left(\mathbf{r}, s\right) \tag{44}$$

Knowing the values of $\tilde{\bar{p}}_i^{\alpha}\left(\mathbf{r}\right)$ and $\tilde{M}_{ij}^{\alpha}\left(\mathbf{r}\right)$, we can sample the conformation statistics of entangled strands under flow on the Kuhn scale, which has now been modified to account for the interaction with the self-consistent potential. Using the same known values of $\tilde{\bar{p}}_i^{\alpha}\left(\mathbf{r}\right)$ and $\tilde{M}_{ij}^{\alpha}\left(\mathbf{r}\right)$ we can also reverse now the use of the Green-Kubo relations derived in the previous section, to determine the six independent components of

$$\tilde{\mathbf{A}}^{\alpha}\left(\mathbf{r}\right) = \frac{\widetilde{\overline{\left\langle \mathbf{RR} \right\rangle_{\mathbf{r}}^{\alpha}}}}{\left(N_{\mathrm{e}}^{\alpha} - 1\right) a^2}, \tag{45}$$

where

$$\widetilde{\overline{\langle \mathbf{RR} \rangle}}_{\mathbf{r}}^{\alpha} = \left(N_{\mathrm{e}}^{\alpha} - 1\right) \widetilde{\overline{\langle \mathbf{d}(1)\mathbf{d}(1) \rangle}}_{\mathbf{r}}^{\alpha} + 2\sum_{s=2}^{N_{\mathrm{e}}-1} \left(N_{\mathrm{e}}^{\alpha} - s\right) \widetilde{\overline{\langle \mathbf{d}(1)\mathbf{d}(s) \rangle}}_{\mathbf{r}}^{\alpha} . \tag{46}$$

is the second moment of the end-to-end distance of an entangled strand of type $\alpha$ under flow, centered at site $\mathbf{r}$, modified to account for interaction with the self-consistent potential. Here the renormalized autocorrelations $\widetilde{\overline{\langle \mathbf{d}(1)\mathbf{d}(s) \rangle}}_{\mathbf{r}}^{\alpha}$ are calculated as in Eqs (13)-(14), but using the renormalized scattering probabilities $\tilde{M}_{ij}^{\alpha}(\mathbf{r})$ and orientation probabilities for interacting strands (Eq. (42)), rather than the “bare” scattering probabilities $M_{ij}(\mathbf{r})$ for ideal (non-interacting) strands, and substituting $\tilde{\bar{p}}_i^{\alpha}(\mathbf{r})$ instead of $\bar{p}_i$.

In the equations above, $\tilde{n}^{\alpha}$ enters as multiplicative renormalization factor transforming the parameters (orientational and scattering probabilities) controlling the persistent random walk sampling the strands conformations on the fcc lattice. In an inhomogeneous fluid of entangled chains, the statistical weight $\tilde{n}^{\alpha}$ of free segments of type $\alpha$ exhibits steep gradients across interfaces between macro- and micro-phase separated domains and next to walls. Therefore, if the point $\mathbf{r}$ is in an interfacial region, $\tilde{\mathbf{A}}^{\alpha}(\mathbf{r})$, the second moment of the end-to end distance of an interacting strand of type $\alpha$, and its eigenvalues $\tilde{\lambda}_i^{\alpha}$ and spatial extensions $\sqrt{\tilde{\lambda}_i^{\alpha}}$, may be quite different from $\mathbf{A}(\mathbf{r})$, $\lambda_i$ and $\sqrt{\lambda_i}$, the analogous quantities for an ideal strand under *the same flow*. Similar differences between interacting and ideal unentangled chains at the interface of a sheared polymer blend were studied quantitatively in Ref. [12] using the DSCF theory for unentangled polymer fluids.

**B. Two-Scale/Two-Mode Sime Propagation a Self-Consistent Field**

We are ready now to consider the conformations statistics of an entangled homopolymer chain in a nonuniform flow, composed of $N^{\alpha}$ freely jointed Kuhn segments of species $\alpha$, interacting with a self-consistent potential that is a function of segmental volume fractions $\phi^{\alpha}(\mathbf{r})$, $\phi^{\beta}(\mathbf{r})$, …, of homopolymer chains of species $\alpha$, $\beta$, …. For this purpose, we will use the notation for labeling segments in the 2S2M model of entangled chains under flow, introduced in subsection II.B. See Fig. 1 and the accompanying text to refresh the assumptions and approximations used in defining this model. We will use the persistent random walk *in a self-consistent field* on a Kuhn-scale fcc lattice to model the conformation of each entangled strand, as defined in the previous subsection. We will use the Wiener (uncorrelated) random walk *in a self-consistent field* on a Kuhn-scale fcc lattice to model the conformations of the dangling strands, as was done for unentangled chains in Ref. [1].

We define the propagator $\tilde{G}_i^{\alpha}(k,s,\mathbf{r})$ as the statistical weight to find a terminal segment at site $\mathbf{r}$ of a homopolymer chain of jointed Kuhn segments of type $\alpha$, starting with a dangling strand of $N_{\mathrm{d}}^{\alpha}$ Kuhn segments , followed first by $k$ entanglement strands of $N_{\mathrm{e}}^{\alpha}$, and then by $s$ Kuhn segments in the terminal entanglement strand, using a recursive sime propagation scheme on two scales. On the Kuhn scale, it propagates from $s$ to $s+1$ within the entangled strand labeled $k$

$$\tilde{G}_i^{\alpha}(k,s+1,\mathbf{r}) = \sum_{j\in I_{12}} \tilde{\bar{M}}_{ij}^{\alpha}(\mathbf{r}-\mathbf{a}_i)\tilde{G}_j^{\alpha}(k,s,\mathbf{r}-\mathbf{a}_i), \tag{47}$$

where $1 \le s \le N_{\mathrm{e}}^{\alpha} - 1$ and $2 \le k \le n_{\mathrm{e}}^{\alpha} + 1$. The initial condition for $s$-propagation within the entangled strands $3 \le k \le n_{\mathrm{e}}^{\alpha} + 1$ is set by $k$-propagation on the PP segment scale,

$$\tilde{G}_i^{\alpha}\left(k+1,1,\mathbf{r}\right)=\tilde{n}\left(\mathbf{r}\right)\overline{p}_i^{\alpha}\left(\mathbf{r}-\sqrt{\tilde{\lambda}_1\left(\mathbf{r}\right)}\mathbf{a}_i\right)\tilde{G}^{\alpha}\left(k,1,\mathbf{r}-\sqrt{\tilde{\lambda}_1\left(\mathbf{r}\right)}\mathbf{a}_i\right). \tag{48}$$

where

$$\tilde{G}^{\alpha}\left(k,s,\mathbf{r}\right)=\sum_{i\in I_{12}}\tilde{G}_i^{\alpha}\left(k,s,\mathbf{r}\right) \tag{49}$$

is the total statistical weight to find the Kuhn segment $s$ belonging to strand $k$ at position $\mathbf{r}$ (regardless of the orientation of the link to the preceding Kuhn segment).

According to Eqs. (48)-(49), the conformations of the PP are sampled by an anisotropic Wiener random walk in a self-consistent potential. It has a local step size $\sqrt{\tilde{\lambda}_1\left(\mathbf{r}\right)}a$, the steps are restricted along the fcc lattice directions $\left\{\mathbf{a}_i\right\}_{i\in I_{12}}$, and the stepping probabilities along these directions, renormalized by the self-consistent potential, are $\tilde{n}\left(\mathbf{r}\right)\overline{p}_i^{\alpha}\left(\mathbf{r}-\sqrt{\tilde{\lambda}_1\left(\mathbf{r}\right)}\mathbf{a}_i\right)$. Note that for a lattice site $\mathbf{r}$, the points $\mathbf{r}-\sqrt{\tilde{\lambda}_1\left(\mathbf{r}\right)}\mathbf{a}_i$ do not necessarily belong to the lattice. However, it is not difficult to evaluate the functions $\overline{p}_i^{\alpha}\left(\mathbf{r}-\sqrt{\tilde{\lambda}_1\left(\mathbf{r}\right)}\mathbf{a}_i\right)$ and $\tilde{G}_j^{\alpha}\left(k,1,\mathbf{r}-\sqrt{\tilde{\lambda}_1\left(\mathbf{r}\right)}\mathbf{a}_i\right)$ at these intermediate points off-lattice, by interpolation from their known values at the lattice points.

Since the first segment ($s=1$) in the first entangled strand ($k=2$) is identified with the last segment ($s=N_{\mathrm{d}}^{\alpha}$), in the first dangling strand ($k=2$), we have

$$\tilde{G}_i^{\alpha}\left(2,1,\mathbf{r}\right)=\overline{p}_i^{\alpha}\tilde{G}^{\alpha}\left(1,N_{\mathrm{d}}^{\alpha},\mathbf{r}\right). \tag{50}$$

This provides an initial condition for $s$-propagation within the first entangled strand, labeled $k=2$. Within the first dangling strand, labeled $k=1$, $s$-propagation on the Kuhn scale is defined as

$$\tilde{G}_i^{\alpha}\left(1,s+1,\mathbf{r}\right)=\tilde{\pi}_i\left(\mathbf{r}-\mathbf{a}_i\right)\tilde{G}^{\alpha}\left(1,s,\mathbf{r}-\mathbf{a}_i\right), \tag{51}$$

where

$$\tilde{\pi}_i\left(\mathbf{r}-\mathbf{a}_i\right)=\tilde{n}^\alpha\left(\mathbf{r}\right)\pi_i\left(\mathbf{r}-\mathbf{a}_i\right) \tag{52}$$

is the renormalized anisotropic probability to step along direction $\mathbf{a}_i$ of the Weiner random walk on the fcc lattice that samples the conformations of the first dangling strand with Kuhn scale step size $a$. Note the similarity between Eqs. (48) and (51): they both correspond to anisotropic Weiner random walks, but on different scale. The reader is cautioned that the notation $\tilde{\pi}_i^\alpha$, used here for the anisotropic stepping probability in the Wiener random walk sampling the conformations of dangling strands on the Kuhn scale, is different from the letter used for the analogous quantity in the case of unentangled chains in Ref. [1], to avoid confusion with the notation $\tilde{\lambda}_i^\alpha$ for the eigenvalues of $\tilde{\mathbf{A}}^\alpha$ in this paper.

The statistical weight of the first Kuhn segment in the first dangling strand is identified with the statistical weight for finding a free Kuhn segment at that site,

$$\tilde{G}^\alpha\left(1,1,\mathbf{r}\right)=\tilde{n}^\alpha\left(\mathbf{r}\right), \tag{53}$$

thus providing an initial condition for $s$-propagation within the first dangling strand ($k=1$). Since we identify the last Kuhn segment ($s=N_{\mathrm{e}}^\alpha$) in the last entangled strand ($k=n_{\mathrm{e}}+1$) with the first Kuhn segment ($s=1$) in the last dangling strand ($k=n_{\mathrm{e}}+2$), we get the following initial condition for s-propagation within the last dangling strand:

$$\tilde{G}^\alpha\left(n_{\mathrm{e}}^\alpha+2,1,\mathbf{r}\right)=\sum_{i\in I_{12}}\tilde{G}_i^\alpha\left(n_{\mathrm{e}}^\alpha+1,N_{\mathrm{e}}^\alpha,\mathbf{r}\right). \tag{54}$$

The volume fractions $\phi_{\mathbf{r}}^{\alpha}$ of *connected* $\alpha$ -type Kuhn segments occupying a lattice site $\mathbf{r}$ at a given time are calculated from known values of $P_{\mathbf{r}}^{\alpha}(s)$ at that time, as follows [10],

$$\phi_{\mathrm{r}}^{\alpha}(\mathbf{r}) = C\left[2\sum_{s=1}^{N_{\mathrm{d}}^{\alpha}}\frac{\tilde{G}^{\alpha}(1,s,\mathbf{r})\tilde{G}^{\alpha}(n_{\mathrm{e}}+2,N_{\mathrm{d}}^{\alpha}-s+1,\mathbf{r})}{\tilde{n}^{\alpha}(\mathbf{r})}\right.$$
$$\left.+\sum_{k=2}^{n_{\mathrm{e}}+1}\sum_{s=1}^{N_{\mathrm{e}}^{\alpha}-1}\frac{\tilde{G}^{\alpha}(k,s,\mathbf{r})\tilde{G}^{\alpha}(n_{\mathrm{e}}+2-k+1,N_{\mathrm{e}}^{\alpha}-s+1,\mathbf{r})}{\tilde{n}^{\alpha}(\mathbf{r})}\right], \quad (55)$$

According to Eq. (55), first the statistical weight for having an $\alpha$ -type Kuhn segment at the lattice site $\mathbf{r}$ while belonging to the $s$ th segment within the $k$ th strand of an entangled chain consisting of $N^{\alpha}$ Kuhn segments is calculated. It is expressed as a product of the statistical weights for two sub-chains starting on the opposite ends of the entangled chain and terminating at the same $s$ th segment within the $k$ th strand, which is then divided by $\tilde{n}^{\alpha}(\mathbf{r})$ to compensate for double counting the statistical weight to place a free segment there. To get the volume fraction $\phi_{\mathbf{r}}^{\alpha}$ at site $\mathbf{r}$ , these statistical weights are then summed over all possible sime values of the common terminal segment along the chain, and normalized. The normalization constant is

$$C_{\alpha} = \frac{\bar{\phi}^{\alpha} n_{\mathrm{s}}}{N^{\alpha}\sum_{\mathbf{r}}\tilde{G}^{\alpha}(n_{\mathrm{e}}+2,N_{\mathrm{d}}^{\alpha},\mathbf{r})}, \quad (56)$$

where $n_{\mathrm{s}}$ is the total number of lattice sites in the finite lattice slab under consideration, and $\bar{\phi}^{\alpha} = \frac{1}{n_s}\sum_{\mathbf{r}}\phi_{\mathbf{r}}^{\alpha}$ is the mean segmental fraction of species $\alpha$ in this slab.

# V. DISCUSSION

We presented here a two-scale/two-mode model for tracking the conformation statistics of entangled polymer chains across interfaces of inhomogeneous polymer fluids evolving under flow. The chains were partitioned into entangled strands between successive entanglements and two terminal dangling strands. The second moment of the end-to end distance of each type of ideal strand under flow followed a different differential evolution equation in continuous space. The second moment was used to regulate the parameters of random walks on the fcc lattice sampling strand's conformation statistics on the Kuhn scale, by means of a generalized Green-Kubo relation and the Maximum Entropy Principle. The random walk parameters were renormalized in the presence of a self-consistent potential representing short-range segmental interactions. The generalized Green-Kubo relation was then inverted to determine how the second moment of strand's end-to-end distance is modified by the self-consistent potential. This allowed us to devise a two-scale sime propagation scheme for the statistical weights of subchains of the entangled chain. The latter was used to calculate local volume fractions for each chemical type of Kuhn segments in entangled chains, thus determining the self-consistent potential.

This work has been motivated by the observation that stretching, squeezing and rotation of entangled strands under flow must induce correlations between successive steps in the random walk sampling its conformation statistics on the Kuhn scale. Such correlations are neglected by the anisotropic Wiener (uncorrelated) random walks on the Kuhn scale that were used in the DSCF theory of unentangled chains [1]. The diffusive propagation in the DSCF model for entangled polymer fluids in Ref. [16] can also be viewed as a continuum limit of an Wiener random walk with anisotropic stepping probabilities, but with the step size corresponding to the equilibrium length of the PP segments; hence it cannot resolve interfacial structure and dynamics arising from segmental interactions on the Kuhn scale.

We devised a flow-regulated persistent random walk model to account for the correlations between successive steps on the Kuhn scale. Similar models accounting for correlations between orientations of bonds or links along the polymer chains have been used since the dawn of Polymer Physics. Simplest examples include the freely rotating chain model and the rotational isomeric state model for polymer conformations, and a variety of models accounting for the effect of semi-flexibility or rigidity in the polymer backbone on conformation statistics [4,33,34]. However, the orientational correlations in these well-known polymer models are typically controlled by fixed interactions parameters. The orientational correlations in the persistent random walks studied in Refs [19,20] as models of mesoscopic diffusion, heat propagation, and anisotropic light scattering in crystalline lattices are also controlled by fixed parameters.

It follows from Eq. (48) that, in our 2S2M model for interacting entangled chains under flow, the PP segments execute an anisotropic Wiener random walks with steps of variable local size set by $\tilde{\lambda}_1^{\alpha}(\mathbf{r})$. We note that the diffusive propagation in DSCF model for entangled polymer fluids in Ref. [16] can also be viewed as a continuum limit of an Wiener random walk with anisotropic stepping probabilities, though with the step size corresponding to the equilibrium length of the PP segments. However the resolution of the model in Ref. [16] is limited by equilibrium length of the PP segments, and hence it cannot resolve interfacial structure and dynamics arising from segmental interactions on the Kuhn scale, while Kuhn scale resolution is built into our 2S2M model.

The novel feature of our persistent random walk model for ideal entangled strands is that the Kuhn-scale orientational correlations are not fixed by interactions, but are rather regulated by the second moment of strand's end-to end distance that evolves under flow. Here we chose to model its evolution by an approximate differential equation (Eqs. (1)-(3)) originally proposed by Marrucci and Ianniruberto [17], due to its simple, computationally efficient form. This equation couples affine deformation by the flow with entropy-driven relaxation of strand stretching and orientation on distinct time scales. However, recent experiments on elongational flows in entangled melts [35] showed significant deviations from the predictions of Eqs. (1)-(3) for the dependence of the steady-state elongational viscosity on the dimensionless elongation rate $\dot{\varepsilon}\tau_{\mathrm{d}}$, in the intermediate regime $1 < \dot{\varepsilon}\tau_d < \tau_d / \tau_{\mathrm{R}}$. A better agreement with experimental data was obtained in a model that accounted for relaxation of the tube diameter driven by interchain pressure effects [36-38], which were neglected in Eqs. (1)-(3)). This was accomplished by a simple dynamic equation for the tube diameter, combining affine deformation by the flow with entropy-driven relaxation dominated by a nonlinear term that arises due to deviation from equilibrium interchain pressure as the tube diameter is squeezed. If $\dot{\varepsilon} \ll \tau_{\mathrm{R}}^{-1}$, one may assume that Rouse relaxation is instantaneous, enforcing constant density of Kuhn segments along a tube section. It might be possible to account for such effects within our 2S2M model by modifying Eqs. (1)-(3) to include a new relaxation term that depends on a *nonlinear*, rotationally invariant function of the eigenvalues $\lambda_i$ of $\mathbf{A}$, in contrast to the Rouse relaxation term that depends on $\mathrm{tr}\,\mathbf{A} = \lambda_1 + \lambda_2 + \lambda_3$, which is a linear, rotationally invariant function of the eigenvalues.

The present formulation of our 2S2M model makes the simplifying assumption that in an entangled linear homopolymer, the number of $\alpha$-type Kuhn segments is fixed at $N_{\mathrm{e}}^{\alpha}$ for an entangled strand and at $N_{\mathrm{d}}^{\alpha}$ for a dangling strand. In reality, the number of Kuhn segments in a strand is not fixed, but rather fluctuates about these mean values, since Kuhn segments can be exchanged between strands sharing the same entanglement (slip-link). Thus it will be a better approximation to model each strand as an open system of Kuhn segments, characterized by a chemical potential controlling their mean number [39,40], but allowing fluctuations about this mean. The grand-canonical formulation is known to produce better agreement with measured normal stress ratios. The chemical potentials of each strand are all equal along the entangled chain at equilibrium, resulting in constant linear density of Kuhn segments. Nonuniform flows may impose a chemical potential gradient along the entangled chain, thus driving mass transport of Kuhn segments across entanglements (slip-links).

In this paper, we approximated the continuous set of all orientations that the PP segment may assume on a unit sphere by a discrete set of only six orientations allowed on the fcc lattice. This lowers drastically the computational cost, but overdetermines the system of six constraint equations resulting from the Green-Kubo relation, hence giving rise to negative solutions $\left\{\overline{p}_i\left(\mathbf{A}\right)\right\}_{i\in I_6}$ for orientational probabilities as functions of the second moments of strands end-to-end distance (we called this effect *lattice frustration* in Sec. III). One could relieve lattice frustration by systematically enlarging the discrete set of allowed directions $i \in I_n$, recovering the unit sphere in the limit $n \to \infty$. Then the system of six constraint equations resulting from the Green-Kubo relation becomes underdetermined, but using them as constraints while maximizing the orientational entropy still allows determination of $\left\{\bar{p}_i\left(\mathbf{A}\right)\right\}_{i\in I_n}$. However, this is impractical, since it increases the computational cost enormously. The use of extended (signed) probability measures and the associated multistate Markov pseudo-process to generalize the persistent random walk model serves as a book-keeping device for the intermediate stages in the calculations, that allows relief of lattice frustration at no extra computational cost, provided that proper entropy measures are used, and that any probabilities used to evaluate observable expectation values are in the interval $[0,1]$.

There are other occasions in physics, where such extended (signed) probability measures occur naturally under somewhat similar circumstances. The Wigner phase-space distribution [24,41,42] is a prominent example of such an extended probability measure. It assumes negative values in the regions of phase space that violate the Heisenberg uncertainty principle for position and momentum, which plays the same role as the constraints imposed by the Green-Kubo relation on the orientations of the PP segments under shear flow in our case. In general, when extended probabilities are forced to assume real values outside the interval $[0,1]$ by some constraints, this signifies that the states corresponding to these values violate the constraints. This does not diminish the usefulness of the Wigner distribution, though modified entropy measures have to be used in this case as well [43]. There are also many examples of extended probabilities used as a book-keeping device in intermediate calculations in classical physics, and in stochastic and pseudo-stochastic processes giving rise to diffusion equations under certain boundary conditions, and to high-order heat-type equations [24,26,44].

It is interesting to note that the continuous limit of a persistent random walk on a one-dimensional lattice results in the one-dimensional telegrapher's equation, which is a partial differential equation of the hyperbolic type. It differs from the heat/diffusion (parabolic type) equation obtained from the continuous limit of a Wiener random walk by an additional term that is proportional to a second-order partial derivative with respect to time. This additional term arises from the second-order Markov nature of the persistent random walk reflecting correlations between successive time (sime) steps. This additional term adds inertial effects to the propagation, on top of the purely diffusive propagation produced by the terms of the heat/diffusion (parabolic) equation. However, on lattices in higher dimensions, the continuous limit of the persistent random walk results in equations with higher order spatial derivatives, [19,20]. On the other hand, higher-order spatial derivatives are known to produce solutions corresponding to signed probability measures in the case of higher order parabolic [26] and hyperbolic [27] equations.

Our use of extended probabilities followed here a point of view that was most eloquently advocated by Richard Feynman in one of the last papers that he submitted before his death, entitled “Negative probabilities.” Since it was published in a festschrift honoring David Bohm that may not be widely available, we believe that it is apt to conclude our paper with an extensive quotation from that source:

“…If a physical theory for calculating probabilities yields a negative probability for a given situation under certain assumed conditions, we do not conclude the theory is incorrect. Two other possibilities of interpretation exist. One is that the conditions (for example, initial conditions) may not be capable of being realized in the physical world. The other possibility is that the situation for which the probability appears to be negative is not one that can be verified directly. A combination of these two, limitation of verifiability and freedom in initial conditions, may also be a solution to the apparent difficulty….

…Since the result must ultimately have a positive probability, the question may be asked: Why not rearrange the calculation so that the probabilities are positive in all the intermediate states? The same question might be asked of an accountant who subtracts the total disbursements before adding the total receipts. He stands a chance of going through an intermediary negative sum. Why not rearrange the calculation? Why bother? There is nothing mathematically wrong with this method of calculating and it frees the mind to think clearly and simply in a situation otherwise quite complicated. An analysis in terms of various states and conditions may simplify a calculation at the expense of requiring negative probabilities for these states. It is not really much expense…

…we would like to emphasize the idea that negative probabilities in a physical theory does not exclude that theory, providing special conditions are put on what is known or verified. But how are we to find and state these special conditions if we have a new theory of this kind? It is that a situation for which a negative probability is calculated is impossible, not in a sense that the chance for it happening is zero, but rather in the sense that the assumed conditions of preparation or verification are experimentally unattainable." [44]

## FIGURES

**Fig. 1.** A schematic sketch of the two-scale/two-mode (2S2M) model of an entangled chain. On the strand scale, entanglements (stars) are connected by primitive path (PP) segments (straight lines) distributed about a preferred orientation. Ellipsoids denote second moments of strand's end-to-end distance. Identical entangled strands are spanned across the PP segments and stretched from isotropic equilibrium conformation along a major axis parallel to the PP segment, and shrunk along the transverse minor axes. Two dangling strands at different orientations are shown at either end of the entangled chain. Blow-ups display an entangled strand and a dangling strand, shown as freely jointed chains of Kuhn segments on the Kuhn scale. See the text for a detailed description of the two scale $(k,s)$ sime indexing scheme for the Kuhn segments.

**Fig. 2.** Time evolution of $\mathrm{tr}\,\mathbf{A}$ at start-up of simple flows. (a) Elongational flows. Dash-dotted (red online), dashed (green online) and solid (blue online) curves are for dimensionless elongation rates $\dot{\varepsilon}\tau_{\mathrm{d}} = 500$, 100 and 10, respectively. (b) Shear flows. Dash-dotted (red online), dashed (green online) and solid (blue online) curves are for dimensionless shear rates $\dot{\gamma}\tau_{\mathrm{d}} = 1000$, 100 and 0.1, respectively.

**Fig. 3.** (a) Time evolution of $\sqrt{\lambda_i}$, where $\lambda_1 \geq \lambda_2 = \lambda_3$ the three eigenvalues of $\mathbf{A}$, at start-up of simple elongational flows. Curves from left to right correspond to dimensionless elongation rates $\dot{\varepsilon}\tau_{\mathrm{d}} = 500$, 100 and 10, respectively, with $\sqrt{\lambda_1}$ shown as dashed lines (red online), and $\sqrt{\lambda_2} = \sqrt{\lambda_3}$ shown as solid lines (blue online). (b) Time evolution of $\sqrt{\lambda_i}$, where $\lambda_1 \geq \lambda_2 \geq \lambda_3$ the three eigenvalues of $\mathbf{A}$ at start-up of simple shear flows. Curves from left to right correspond to dimensionless shear rates $\dot{\gamma}\tau_{\mathrm{d}} = 1000$,

100 and 10, respectively, with $\sqrt{\lambda_1}$ shown as black dash-dotted lines , $\sqrt{\lambda_2}$ as solid lines (blue online), and $\sqrt{\lambda_3}$ as dashed lines (red online). (c) Time evolution of $\theta$, the angle between the eigenvector associated with the eigenvalue $\lambda_1$ and $\mathbf{e}_1$, at start-up of simple shear flows. Triangles (green online), squares (blue online) and circles (red online) correspond to dimensionless shear rates $\dot{\gamma}\tau_\mathrm{d} = 10$, 100 and 1000, respectively.

**Fig. 4.** Log-log plots of the steady-state values of $\sqrt{\lambda_i}$ as functions of the dimensionless deformation rate for simple flows. (a) Elongational flows. Stars (red online) correspond to $\sqrt{\lambda_1}$ , circles (blue online) to $\sqrt{\lambda_2} = \sqrt{\lambda_3}$ . Asymptotic dashed line at high elongation rate has the slope $-0.500$. (b) Shear flows. Black dots correspond to $\sqrt{\lambda_1}$ , circles (blue online) to $\sqrt{\lambda_2}$ , and stars (red online) to $\sqrt{\lambda_3}$ . Asymptotic dashed lines at high shear rates have identical slopes $-0.315$ for $\sqrt{\lambda_2}$ (blue online) and $\sqrt{\lambda_3}$ (red online).

**Fig. 5.** Schematic picture of the site $\mathbf{r}$ (labeled 0) and its twelve nearest neighbors $\mathbf{r} + \mathbf{a}_i$ on the fcc lattice (labeled $\pm i$). The black sites labeled $\pm 1$, $\pm 2$ and $\pm 3$ are on a triangular lattice layer in the *xy* plane at $z = 0$. The sites labeled 4, 5 and 6 (blue online) are on a triangular lattice layer in the *xy* plane at $z = \sqrt{2/3}a$ . The sites labeled $-4$, $-5$ and $-6$ (red online) are on a triangular lattice layer in the *xy*-plane at $z = -\sqrt{2/3}a$ . The distance between the $xy$ planes is not shown to scale.

**Fig. 6.** Time evolution of orientational probabilities $\{\overline{p}_i\}_{i \in I_6}$ for the PP segments of ideal strands at start-up of simple flows. For $\{\overline{p}_{-i}\}_{i \in I_6}$ use $\overline{p}_{-i} = \overline{p}_i$ . (a) Elongational flow at dimensionless elongation rate $\dot{\varepsilon}\tau_\mathrm{d} = 500$. Dashed curve (blue online) corresponds to $\overline{p}_1(t)$, solid curve (red online) shows overlapping $\overline{p}_i(t)$ for $i = 2,\ldots,6$. (b) Shear flow at dimensionless shear rate $\dot{\gamma}\tau_{\mathrm{d}} = 1000$. The dashed curve (blue online) corresponds to $\overline{p}_1(t)$, the solid curve (red online) to $\overline{p}_2(t)$, the dash-dotted curve (green online) to $\overline{p}_3(t)$, and the dots (magenta online) to the overlapping curves $\overline{p}_4(t) = \overline{p}_5(t) = \overline{p}_6(t)$.

**Fig. 7.** Comparison of different entropy measures for orientational probabilities of the PP segments of ideal entangled strands at the start-up of simple flows, as a function of time. (a) Elongational flow at dimensionless elongations rate $\dot{\varepsilon}\tau_{\mathrm{d}} = 500$. (b) Shear flow at dimensionless shear rate $\dot{\gamma}\tau_{\mathrm{d}} = 1000$. In both (a) and (b), dash-dotted curves (red online) show the second-order Renyi entropy, and the insets show the second-order Tsallis entropy as a dashed line (magenta online) on a different vertical scale. In (a), solid curve (blue online) shows the Shannon entropy curve. Only the real-valued branch of the Shannon entropy curve (circles, blue online) is shown in (b).

**Fig. 8.** Time evolution of the second-order Renyi entropy $S^R_{2\overline{\mathbf{p}}}$ of the orientational probabilities of the PP segments of ideal entangled strands at the start-up of simple flows. (a) Elongational flow. Solid (blue online), dashed (green online) and dash-dotted (red online) curves correspond to dimensionless elongation rates $\dot{\varepsilon}\tau_{\mathrm{d}} = 10$, 100 and 500, respectively. (b) Shear flow. The straight solid line (blue online), and the dash-dotted (green online), dashed (magenta online) and solid (red online) curves correspond to dimensionless shear rates $\dot{\gamma}\tau_{\mathrm{d}} = 0.1$, 10, 100 and 1000, respectively.

**Fig. 9.** Scattering probabilities as functions of $\mathrm{tr}\,\mathbf{A}$ for ideal entangled strands. Transmission probability $T$ is shown as a solid curve (red online), reflection probability

$R$ as a dashed curve (blue online), and the lateral scattering probability as a dash-dotted curve (magenta online).

**Fig. 10.** Time evolution of the scattering probabilities for ideal entangled strands at start-up of simple flows. $T$, $R$ and $L$ are denoted by solid (blue online), dash-dotted (red online) and black dashed lines, respectively. (a) Elongational flows (log-log plots). Curves denoted by $a, b, c$ correspond to dimensionless elongation rates $\dot{\varepsilon}\tau_{\mathrm{d}} = 500$, 100 and 10, respectively (b) Shear flows (semi-log plots). Curves from left to right correspond to dimensionless shear rates $\dot{\gamma}\tau_{\mathrm{d}} = 1000$, 100 and 0.1, respectively. Note that the curves corresponding to $\dot{\varepsilon}\tau_{\mathrm{d}} = 10$ in (a) and to $\dot{\gamma}\tau_{\mathrm{d}} = 0.1$ in (b) are practically indistinguishable from the constant line with the equilibrium value of $T = R = L = \frac{1}{12}$.

**Fig. 11.** Shannon entropy $S_{\mathbf{M}}$ of the scattering probabilities for ideal strands at the start-up of simple flows, as a function of time. (a) Elongational flows. The solid black line, the dashed (blue online) and dash-dotted (red online) curves correspond to dimensionless elongation rates $\dot{\varepsilon}\tau_{\mathrm{d}} = 10$, 100 and 500, respectively. (b) Shear flows. The dash-dotted curve (red online) corresponds to dimensionless shear rate $\dot{\gamma}\tau_{\mathrm{d}} = 1000$, and the two curves corresponding to $\dot{\gamma}\tau_{\mathrm{d}} = 100$ and 0.1 are indistinguishable from the constant line with the equilibrium value $S_{\mathbf{M}} = \ln 12$ (solid black line).

Fig. 1

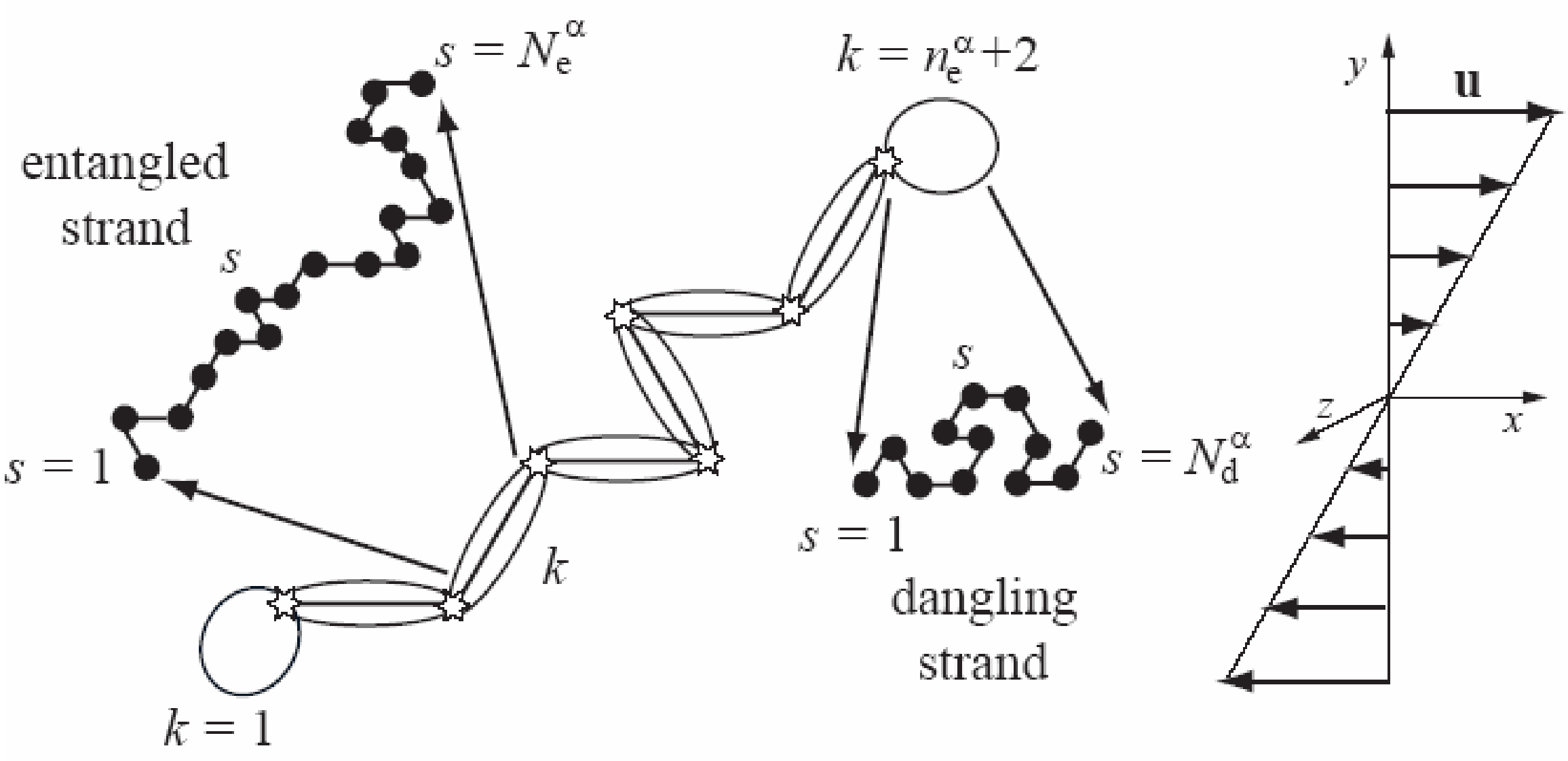

Fig. 2(a)

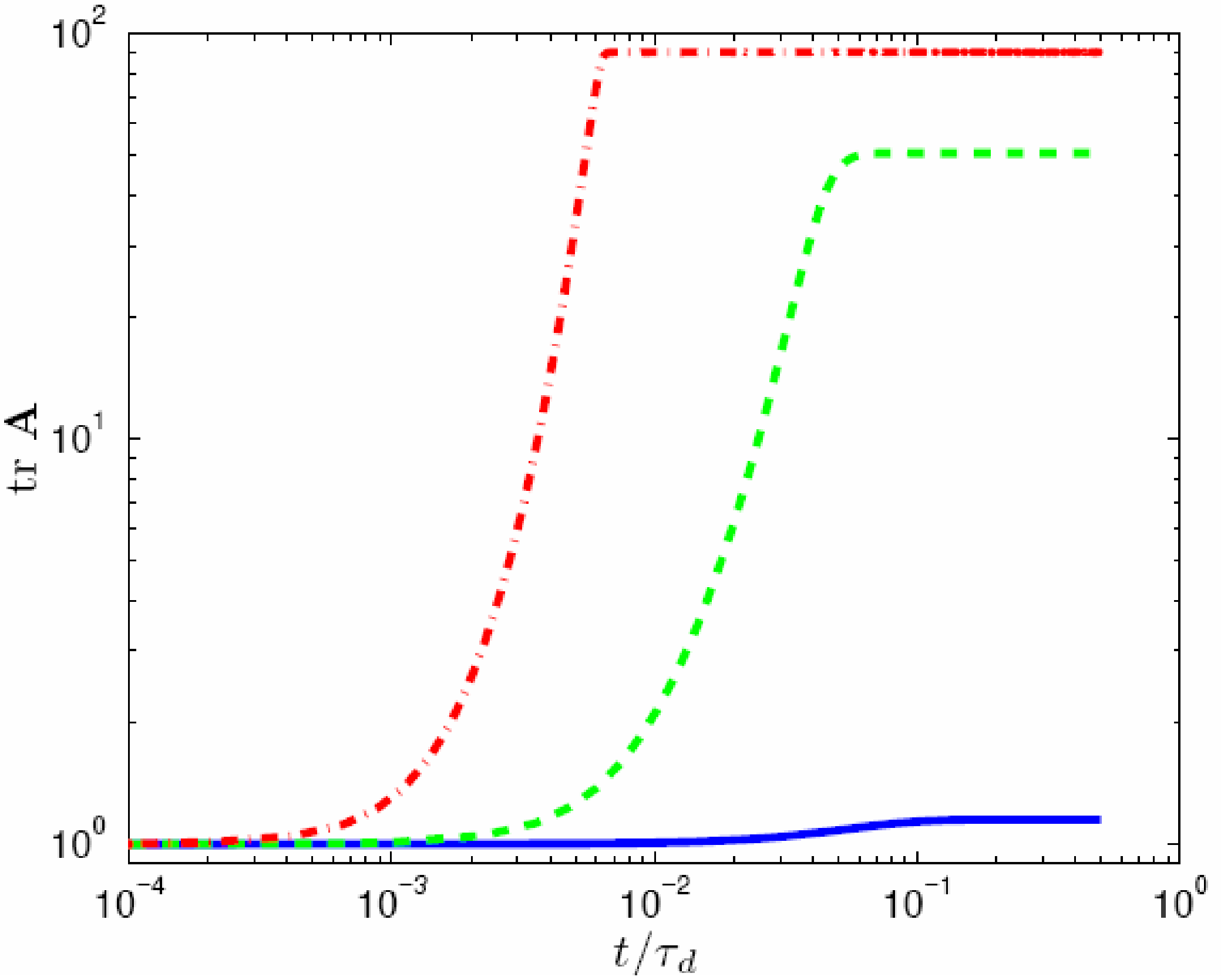

Fig. 2(b)

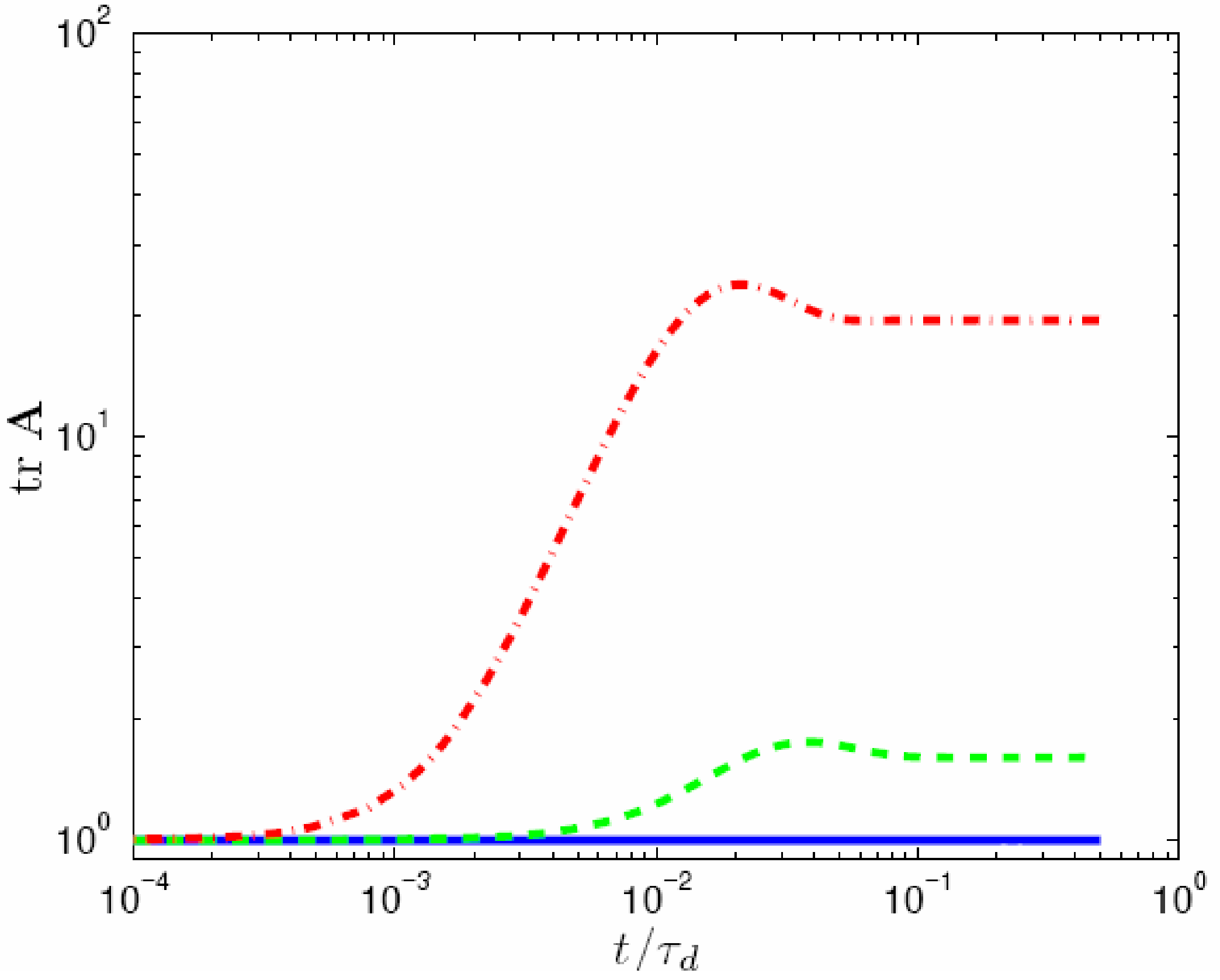

Fig. 3(a)

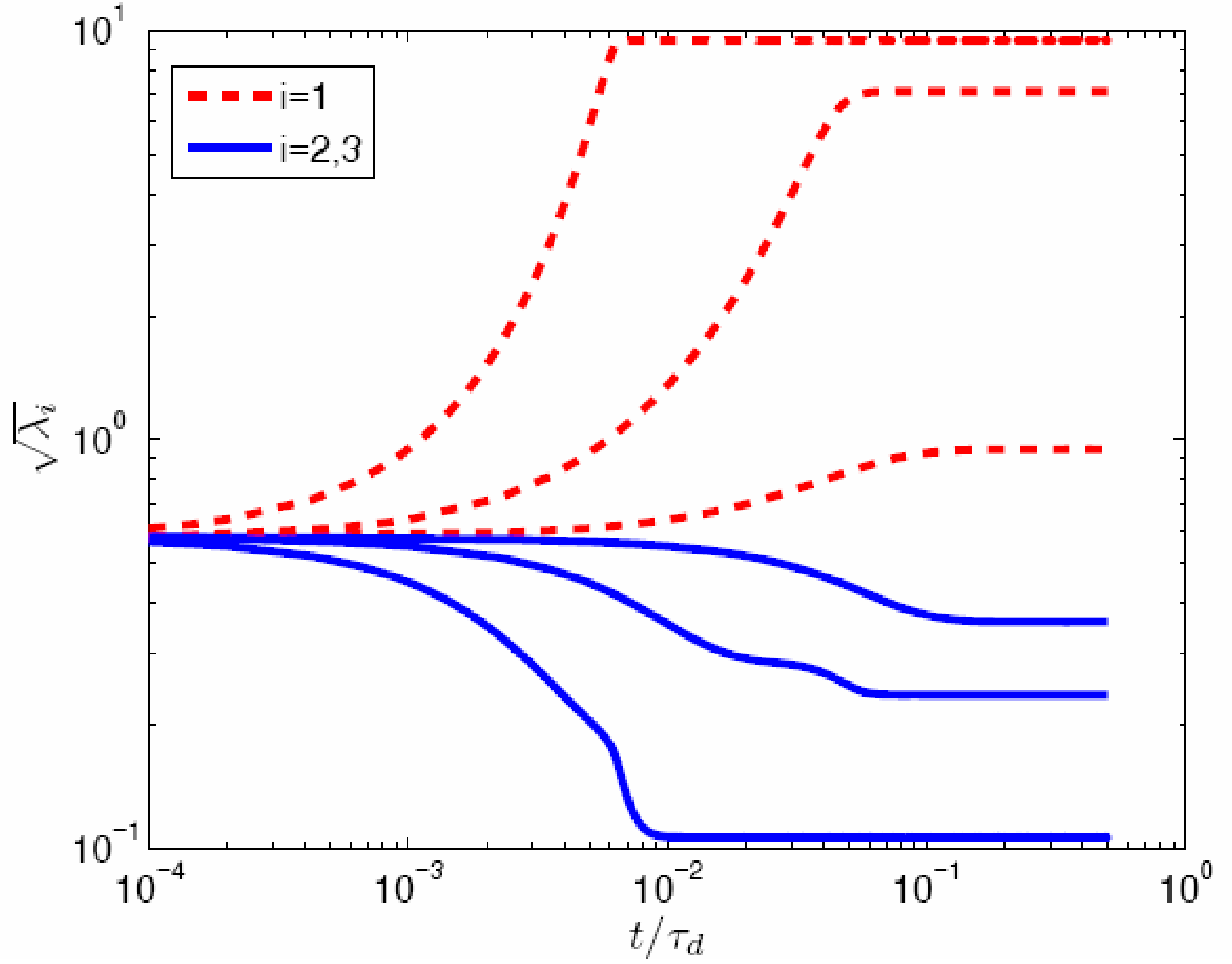

Fig. 3(b)

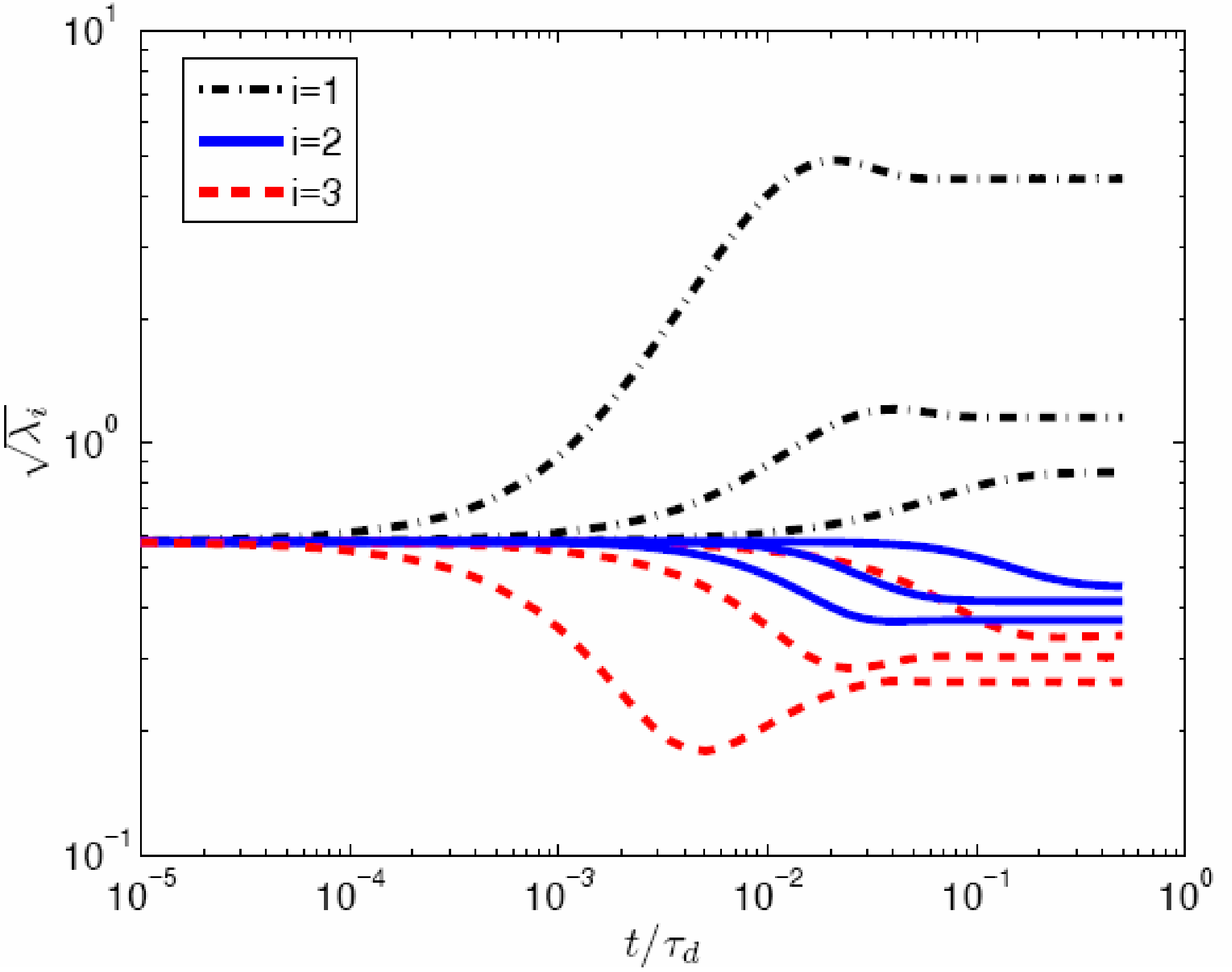

Fig. 3(c)

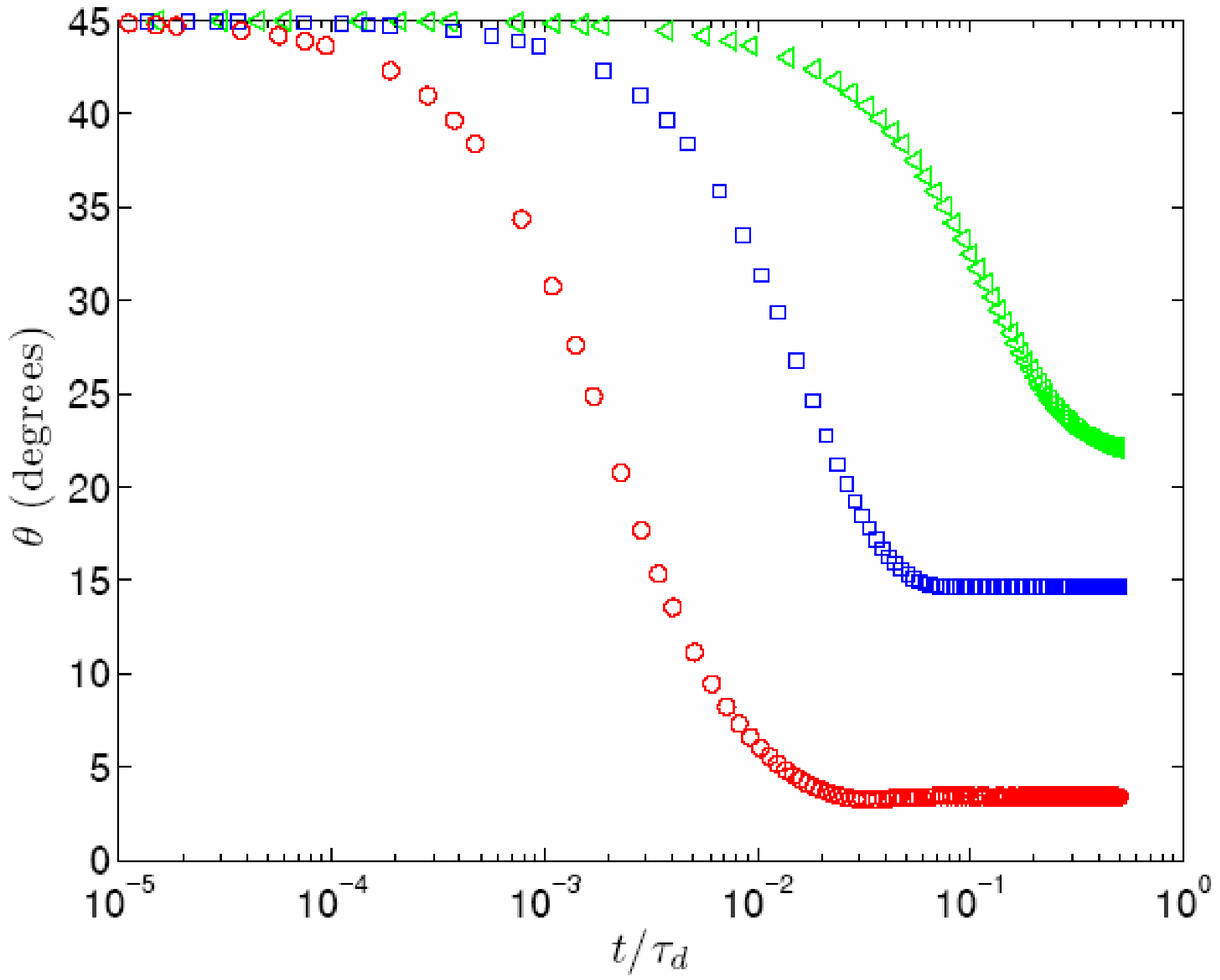

Fig. 4(a)

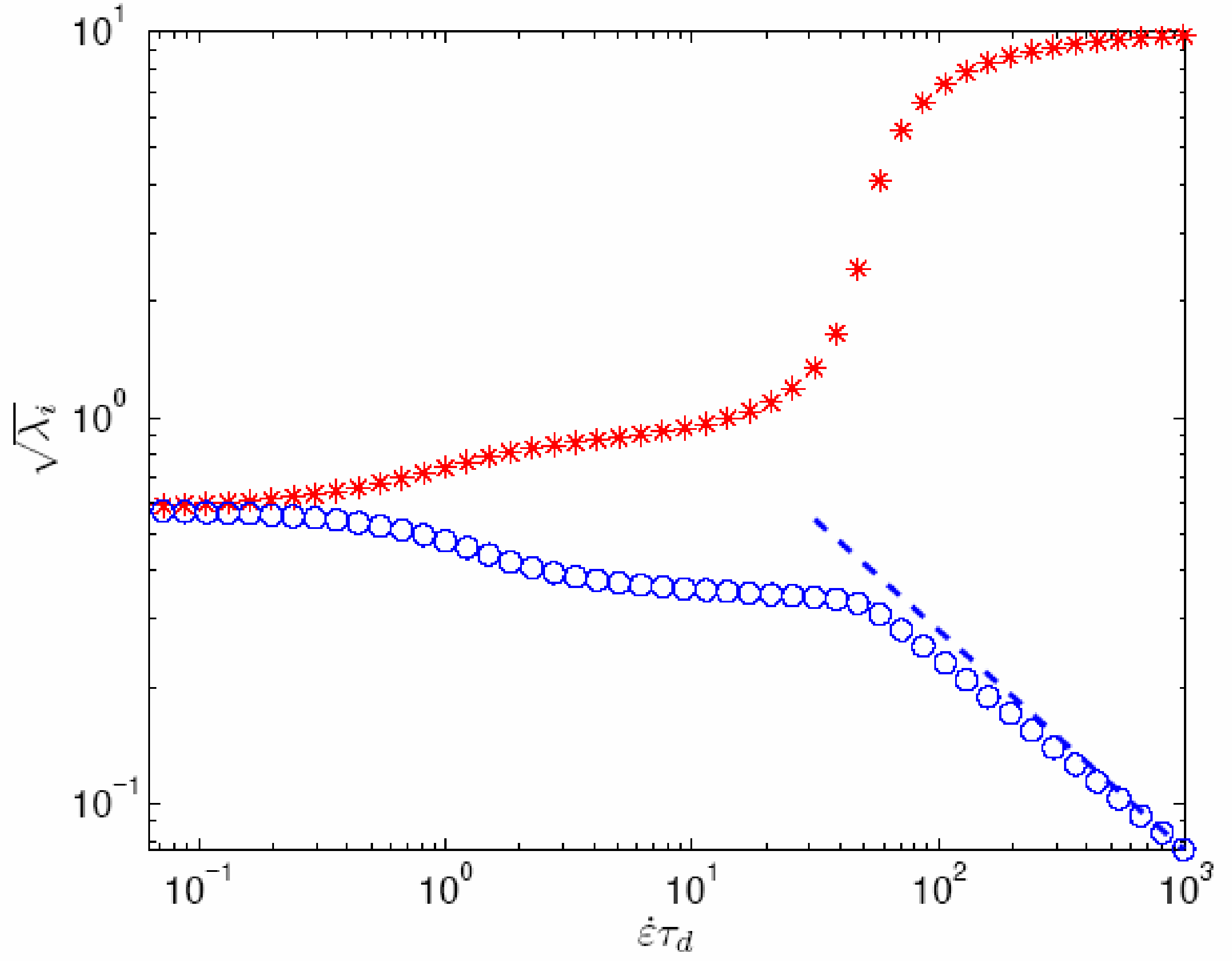

Fig. 4(b)

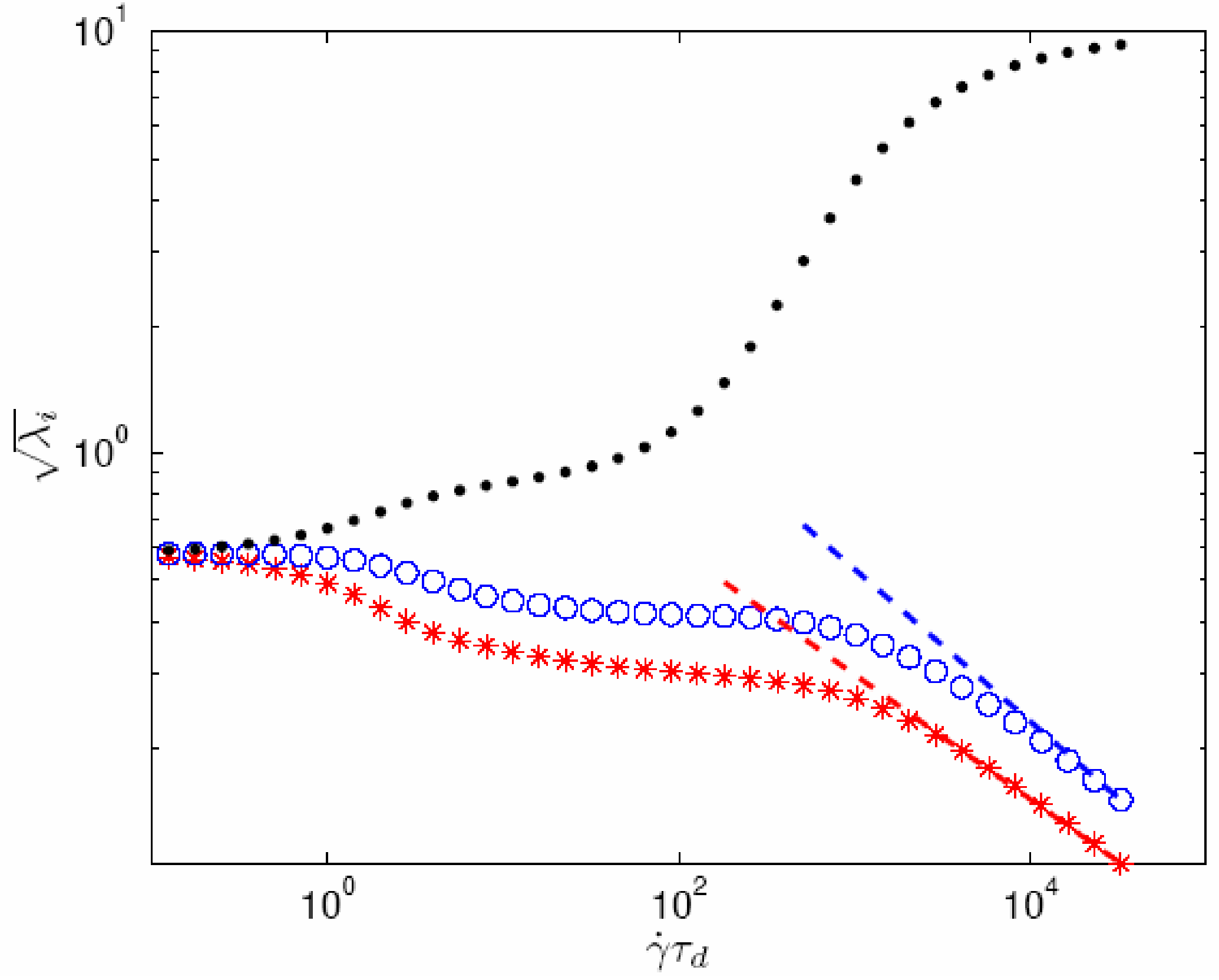

Fig. 5

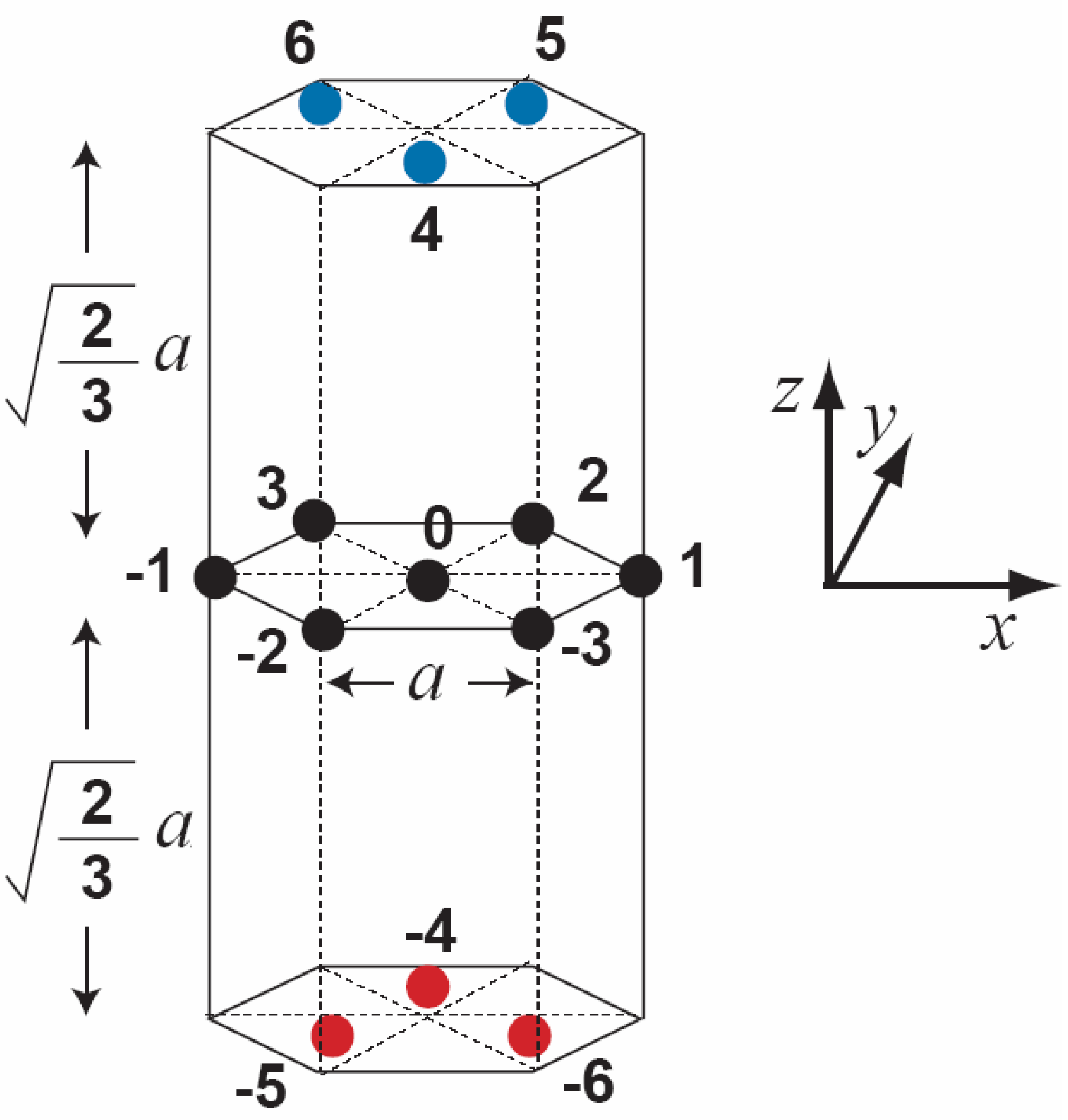

Fig. 6(a)

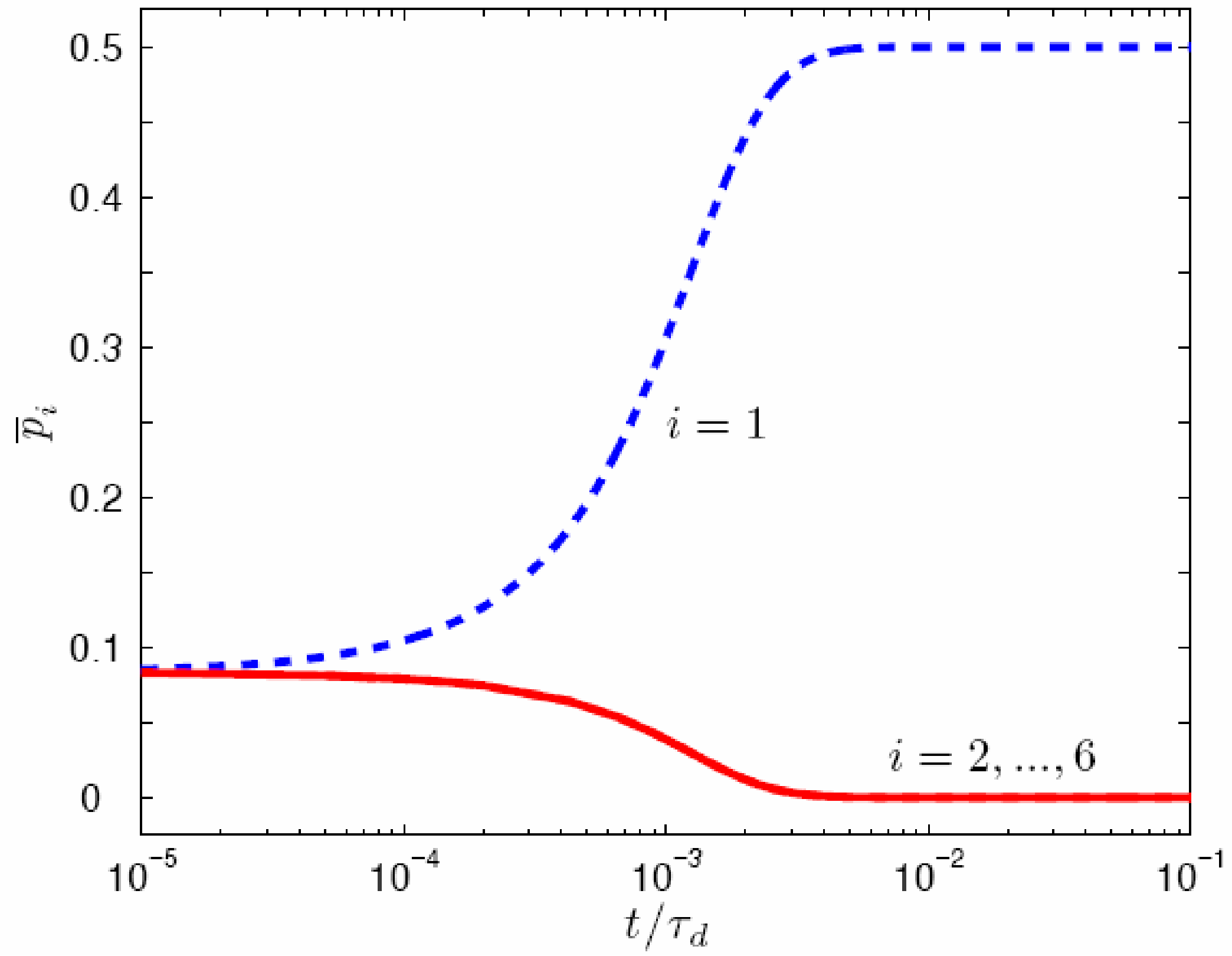

Fig. 6(b)

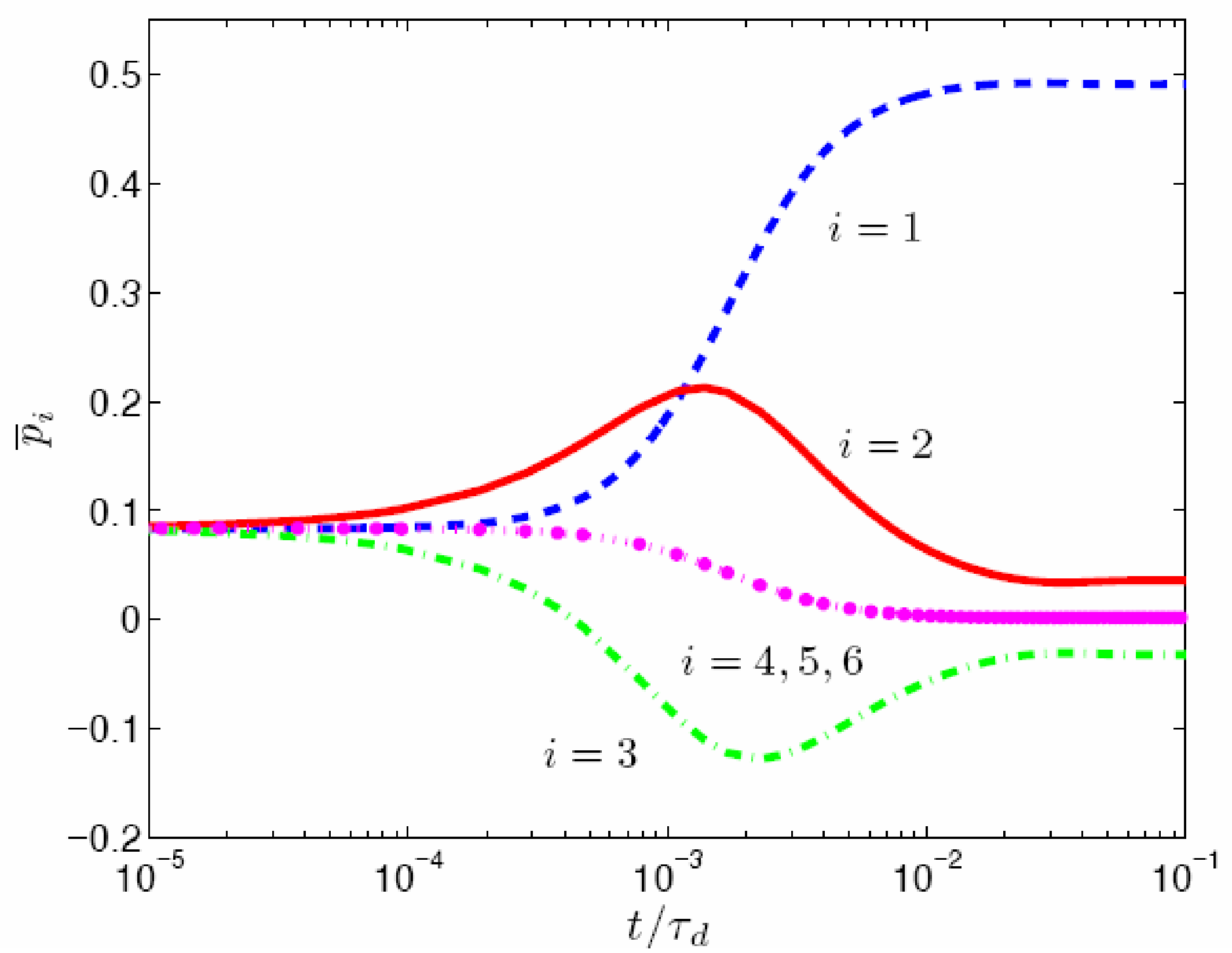

Fig. 7(a)

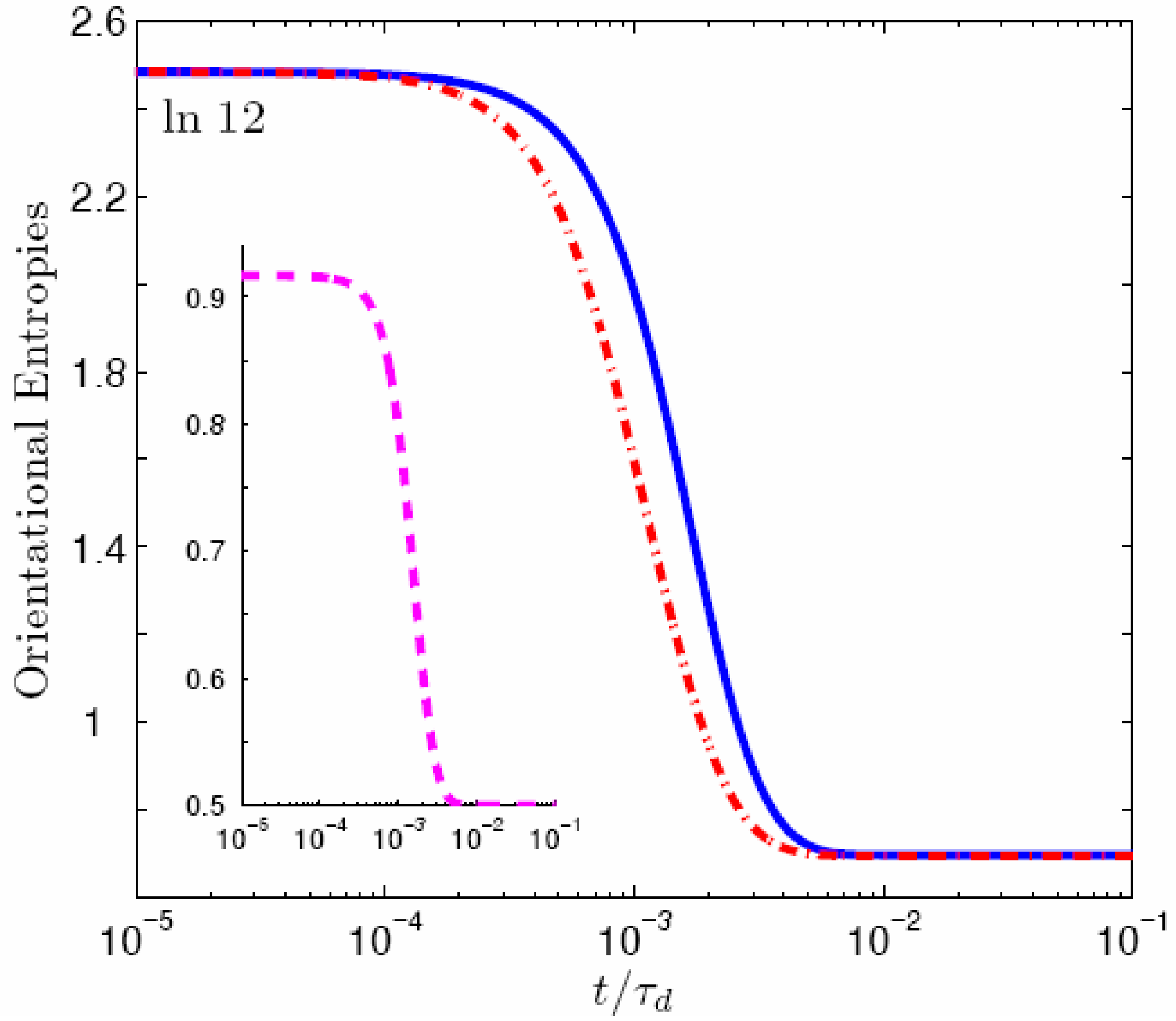

Fig. 7(b)

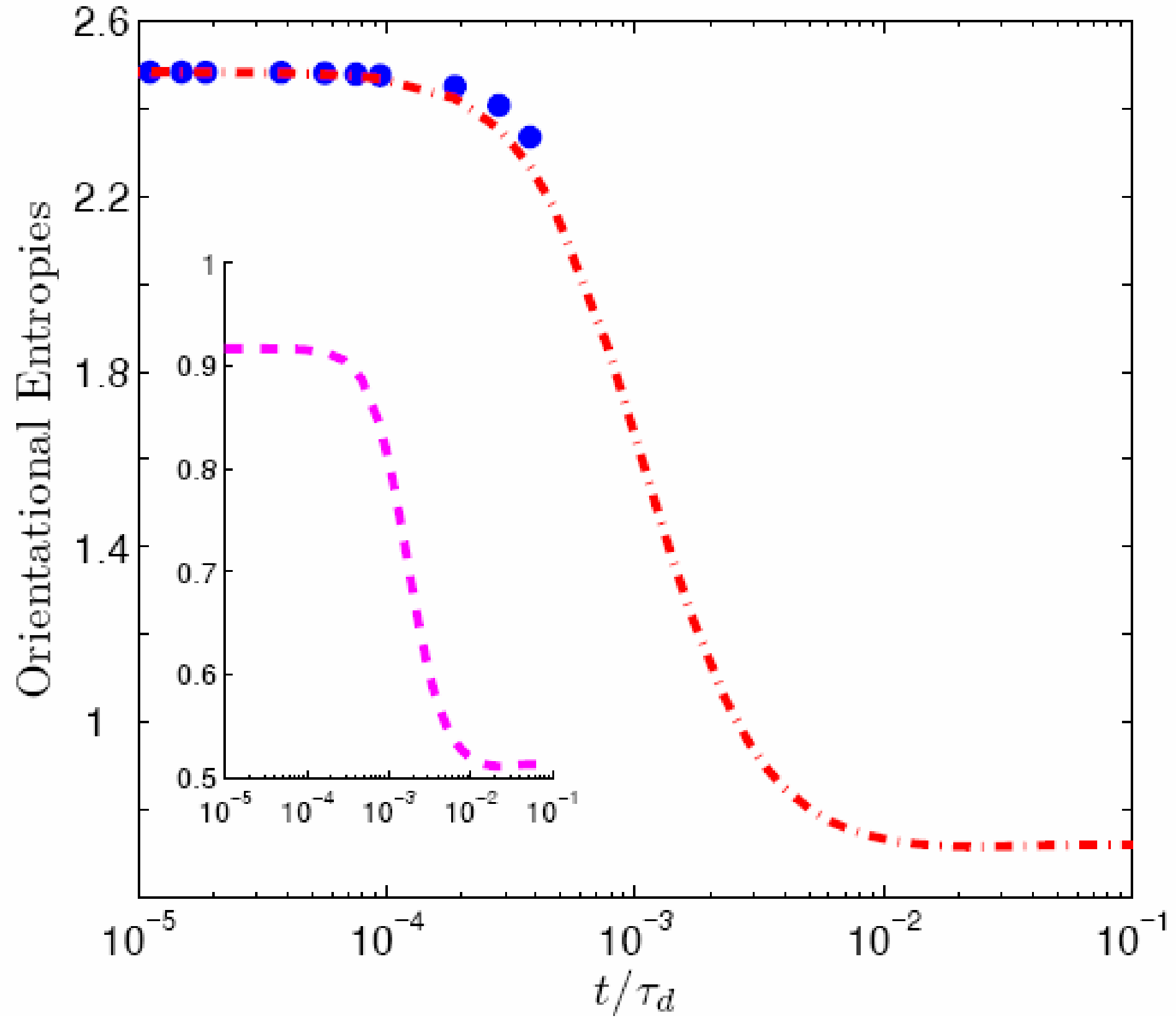

Fig. 8(a)

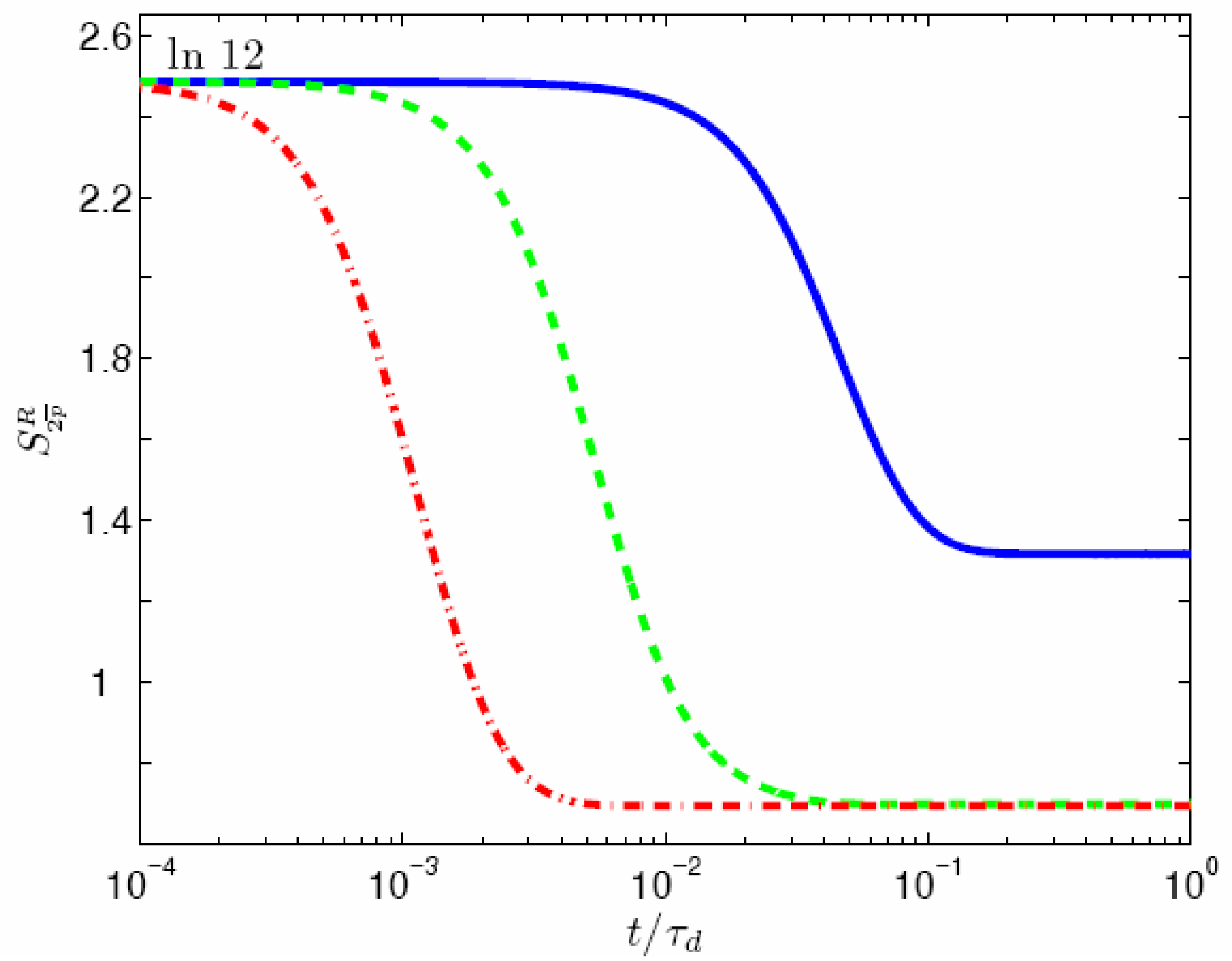

Fig. 8(b)

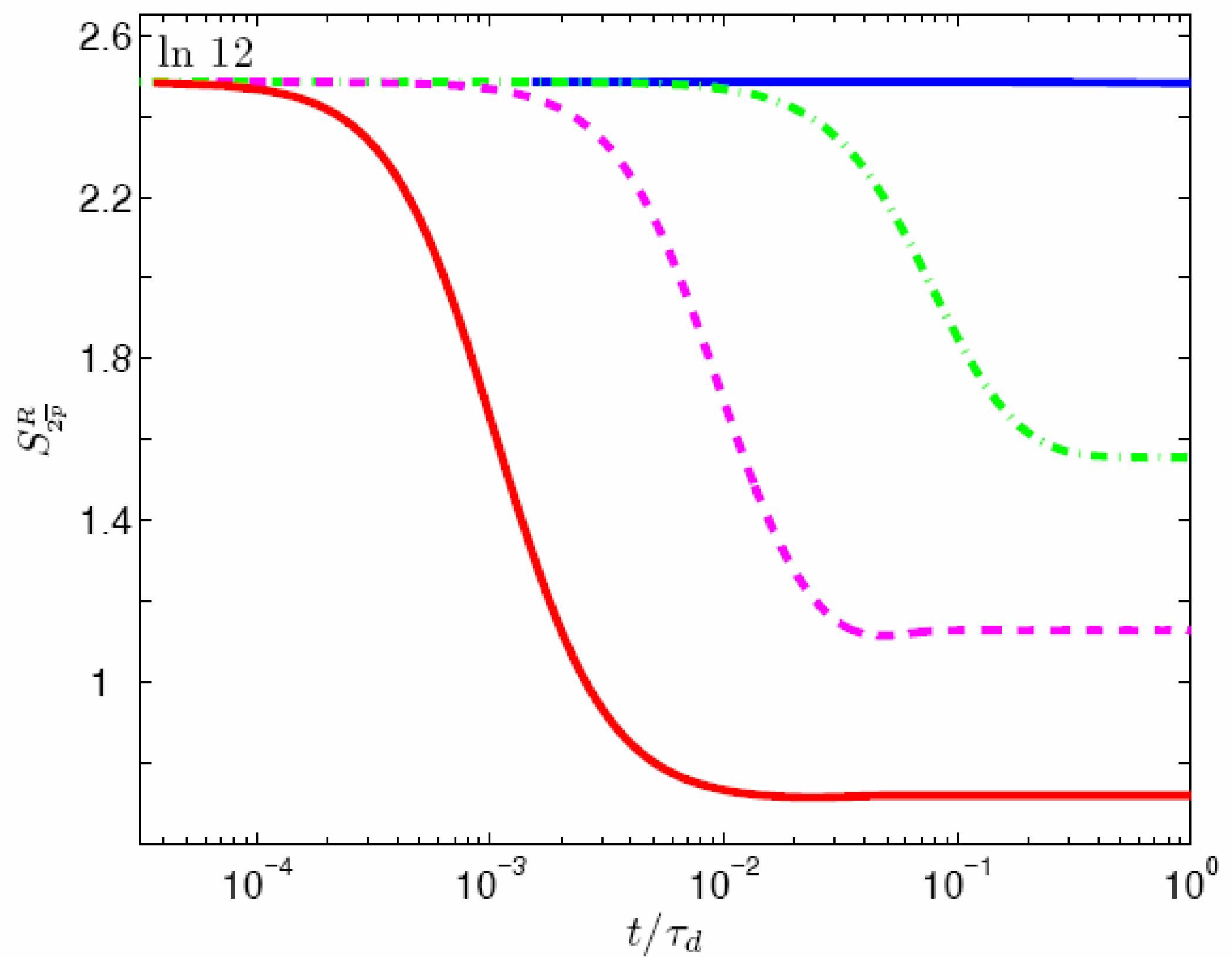

Fig. 9

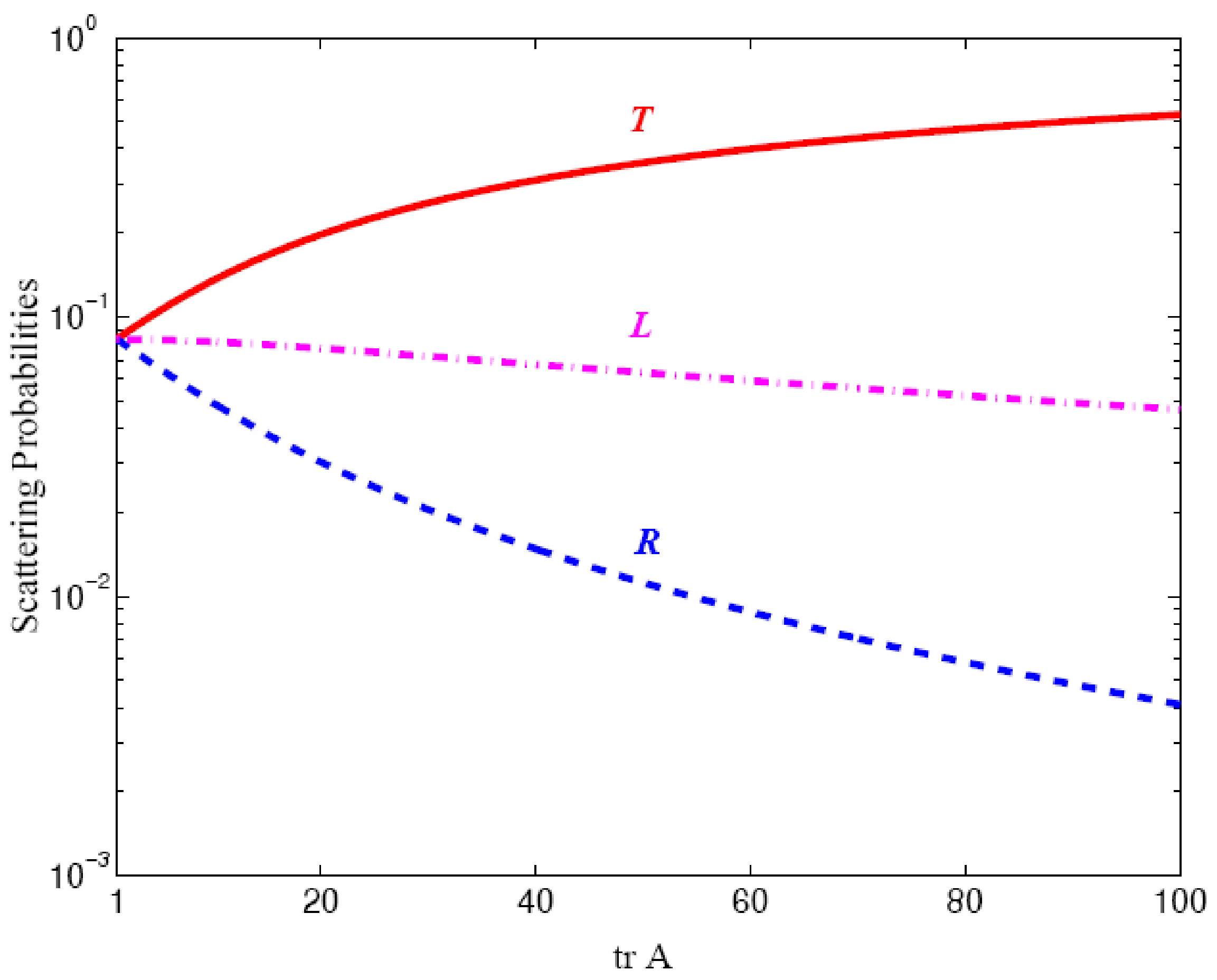

Fig. 10(a)

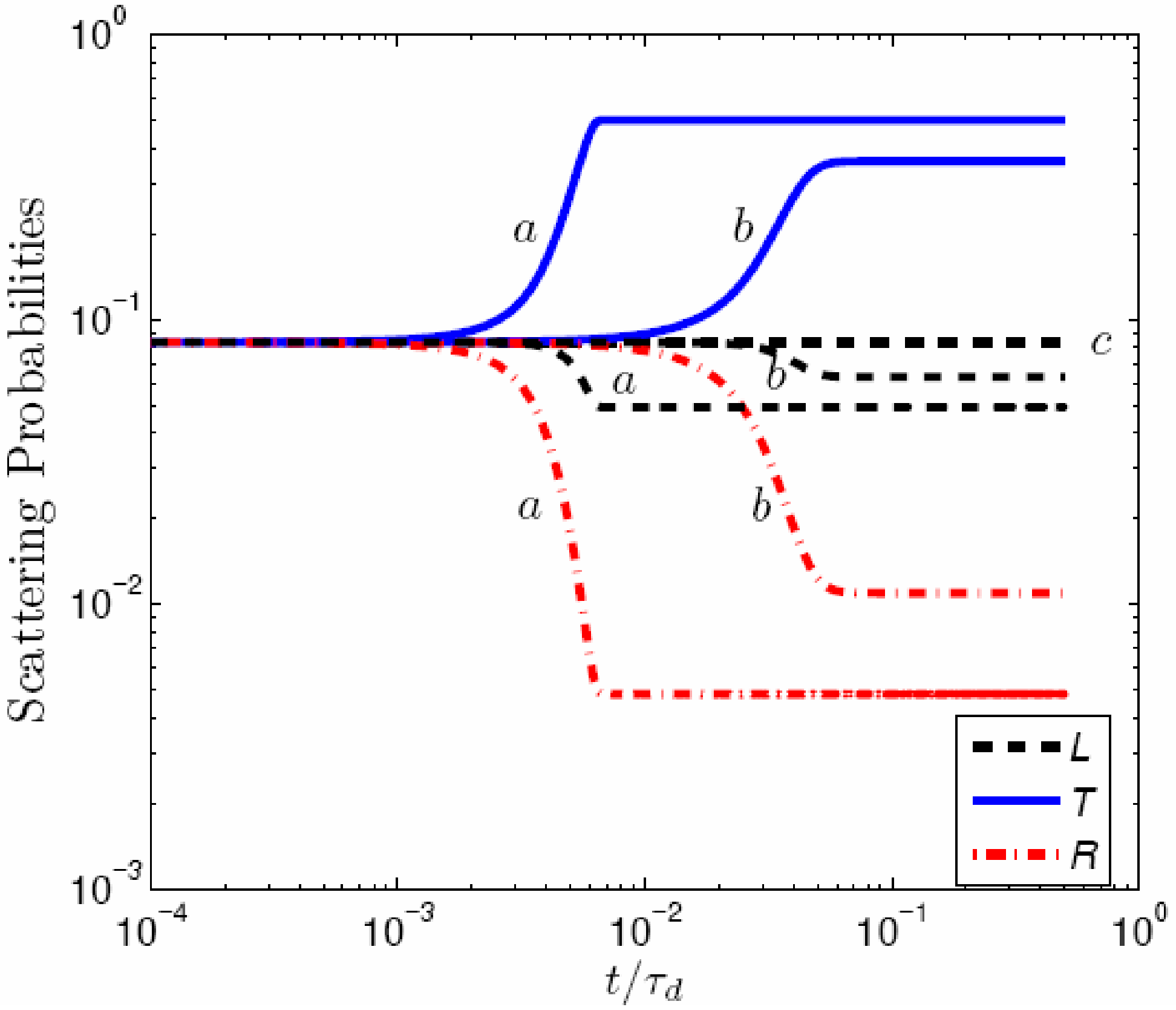

Fig. 10(b)

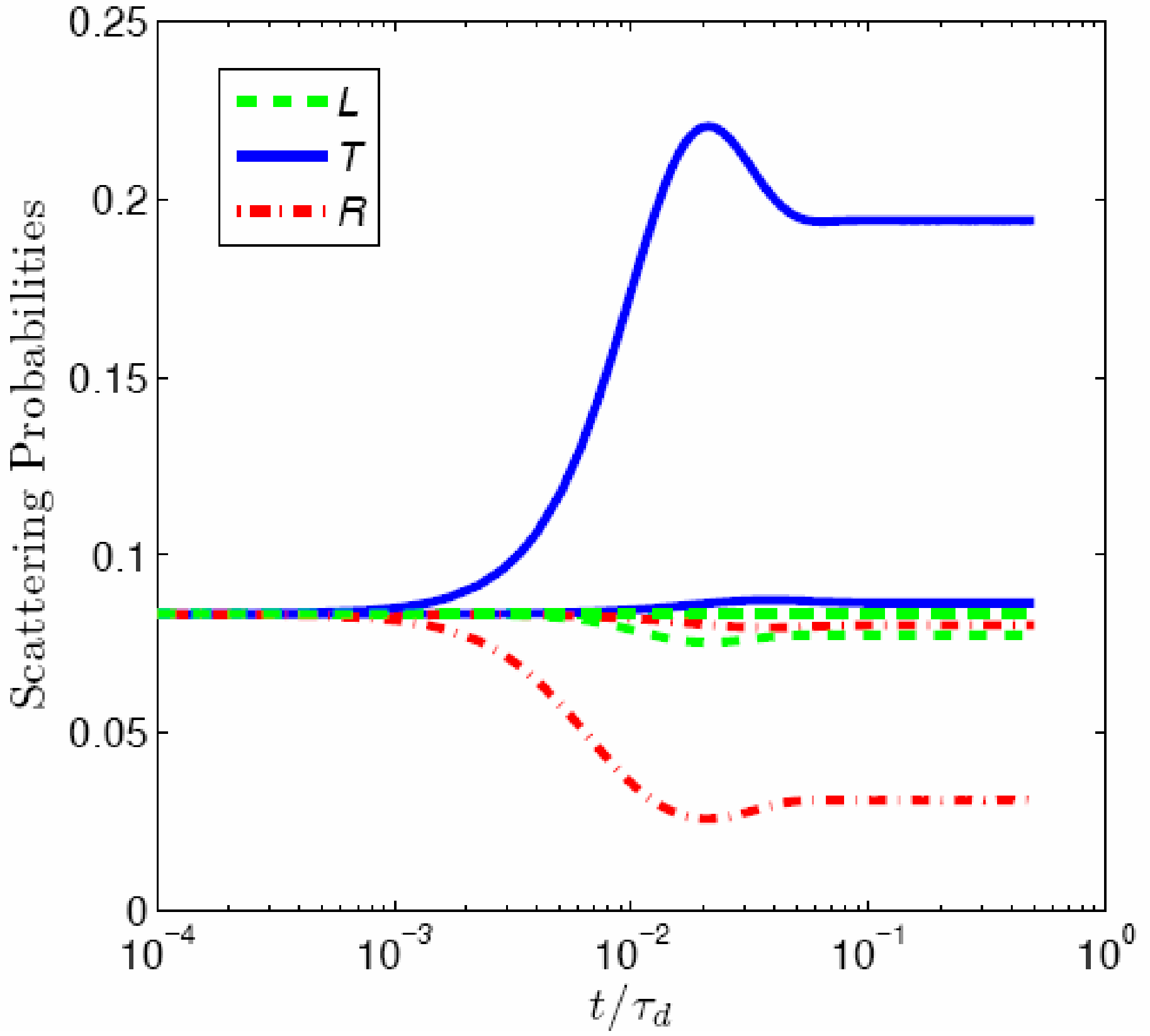

Fig. 11(a)

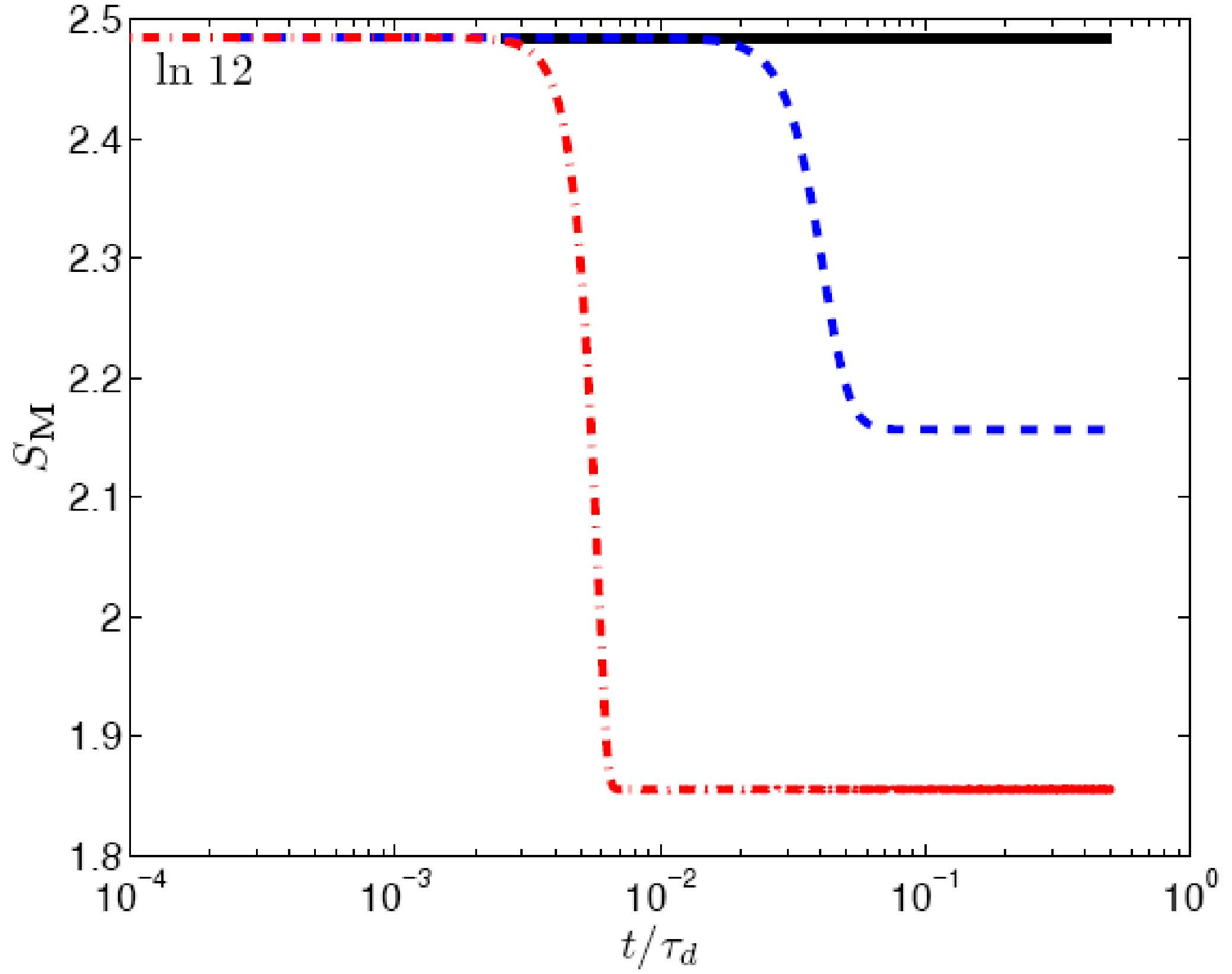

Fig. 11(b)

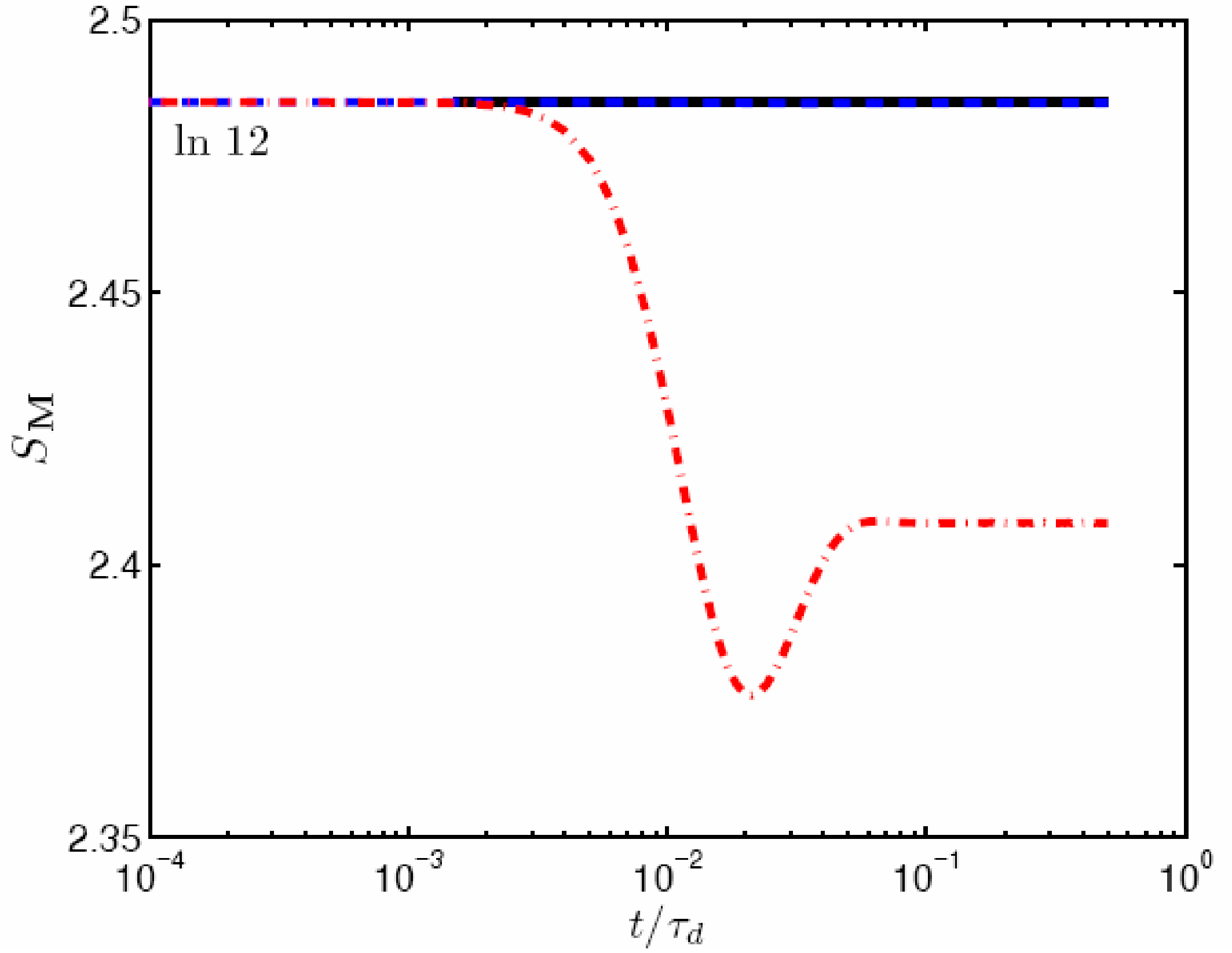